\documentclass[11pt]{article}
\usepackage{geometry}
\usepackage{soul}
\setstcolor{red}
\usepackage{subcaption}
\usepackage{overpic}
\usepackage{pict2e}

\usepackage{amsmath,amssymb}
\allowdisplaybreaks[1]

\usepackage{changepage}

\usepackage[utf8x]{inputenc}

\usepackage{textcomp,marvosym}

\usepackage{cite}
\usepackage{url}

\usepackage{nameref,hyperref}

\usepackage{microtype}
\DisableLigatures[f]{encoding = *, family = * }

\usepackage[table]{xcolor}

\usepackage{array}
\usepackage{placeins}

\newcolumntype{+}{!{\vrule width 2pt}}

\newlength\savedwidth

\usepackage{setspace} 
\usepackage[aboveskip=1pt,labelfont=bf,labelsep=period,singlelinecheck=off]{caption}

\makeatletter
\renewcommand{\@biblabel}[1]{\quad#1.}
\makeatother

\usepackage{lastpage,fancyhdr,graphicx}
\usepackage{epstopdf}

\usepackage{atbegshi,picture}

\AtBeginShipout{\AtBeginShipoutUpperLeft{%
  \put(\dimexpr\paperwidth-5cm\relax,-1.5cm){\makebox[0pt][r]{\framebox{Link to the journal version: \url{https://doi.org/10.1016/j.mbs.2021.108664}}}}%
}}

\begin{document}
\singlespacing

\begin{flushleft}
{\Large
\textbf\newline{An epidemic model for COVID-19 transmission in Argentina: Exploration of the alternating quarantine and massive testing strategies} 
}
\newline
\\
Lautaro Vassallo\textsuperscript{1,2*},
Ignacio A. Perez\textsuperscript{1,2},
Lucila G. Alvarez-Zuzek\textsuperscript{3},
Julián Amaya\textsuperscript{2},
Marcos F. Torres\textsuperscript{1,2},
Lucas D. Valdez\textsuperscript{1,2},
Cristian E. La Rocca\textsuperscript{1,2},
Lidia A. Braunstein\textsuperscript{1,2,4}
\\
\bigskip
\textbf{1} Instituto de Investigaciones Físicas de Mar del Plata (IFIMAR), CONICET - Universidad Nacional de Mar del Plata, 7600 Mar del Plata, Buenos Aires, Argentina
\\
\textbf{2} Departamento de Física, Facultad de Ciencias Exactas y Naturales, Universidad Nacional de Mar del Plata, 7600 Mar del Plata, Buenos Aires, Argentina
\\
\textbf{3} Department of Biology, Georgetown University, Washington, D.C. 20057, United States
\\
\textbf{4} Physics Department, Boston University, Boston, Massachusetts 02215, United States
\\
\bigskip

* lvassallo@mdp.edu.ar

\end{flushleft}

\section*{Abstract}
The COVID-19 pandemic has challenged authorities at different levels of government administration around the globe. When faced with diseases of this severity, it is useful for the authorities to have prediction tools to estimate in advance the impact on the health system {as well as} the human, material, and economic resources that will be necessary. In this paper, we construct an extended Susceptible-Exposed-Infected-Recovered model that incorporates the social structure of Mar del Plata, the 4º most inhabited city in Argentina and head of the Municipality of General Pueyrredón. Moreover, we consider detailed partitions of infected individuals according to the illness severity, as well as data of local health resources, to bring predictions closer to the local reality. Tuning the corresponding epidemic parameters for COVID-19, we study an alternating quarantine strategy: a part of the population can circulate without restrictions at any time, while the rest is equally divided into two groups and goes on successive periods of normal activity and lockdown, each one with a duration of $\tau$ days. {We also} implement a random testing strategy with a threshold over the population. We found that $\tau = 7$ is a good choice for the quarantine strategy since it reduces the infected population {and, conveniently, it suits a weekly schedule}. Focusing on the health system, projecting from the situation as of September 30, we foresee a difficulty to avoid saturation of {the available} ICU, given the extremely low levels of mobility that would be required. In the worst case, our model estimates that four thousand deaths would occur, of which 30\% could be avoided with proper medical attention. Nonetheless, we found that aggressive testing would allow an increase in the percentage of people that can circulate without restrictions, and the medical facilities to deal with the additional critical patients would be relatively low.

\section{Introduction}
In March 2020 the World Health Organization (WHO) officially announced
the COVID-19 (disease caused by the novel SARS-CoV-2 coronavirus) as a
pandemic. The world was about to witness one of the most devastating
pandemics in {the} last decades. In December 2019, a cluster of a novel
pneumonia-like illness was identified in Wuhan (China) and, without the proper
control measures, in 3 months the disease rapidly spread all over the
world. The first stages of the propagation occurred in different
countries of Europe, leading to countless human life losses and a severe
economic crisis.

On March 3, the first COVID-19 patient was confirmed in
Buenos Aires, Argentina. It was an imported case, a 43-year-old man who had
arrived from Milan, Italy. Following international
protocols and recommendations of local experts, educational centers of
all levels were closed, massive shows were suspended and international
frontiers were closed. Decisively, on March 19{,} and in order to restrict the
spread of the new coronavirus, the national government announced a 12
days nationwide lockdown 
\cite{arg_medidas}. These measures {have been progressively extended to the present day} with certain relaxations.
Until the middle of September, the total number of
confirmed cases surpassed $750$ thousand and the number of fatalities was
around $17$ thousand, most of them concentrated in Capital Federal (CABA) and the suburbs, which form the urban
conglomerate {known as Área Metropolitana de Buenos Aires (AMBA)}, inhabited by $37\%$ of
the total population in Argentina. 

Researchers have put a lot of effort in the study of multiple aspects of
COVID-19 pandemic in several countries \cite{vespi_modelling,meyer_outbreak,galvani_travel,colizza_phones,hens_gen_int, juanli,yuan}, at country level and at big
regions with a high concentration of people \cite{mininni}.
{
Amaku {\it et al} \cite{amaku} developed a mathematical model to study two testing strategies to halt the epidemic spread in Brazil. They found that a strategy that considers testing only infected individuals and their immediate contacts has the same effectiveness in reducing the total number of cases as a massive testing strategy but at a much lower cost. Paternina-Caicedo {\it et al} \cite{paternina} studied an intermittent intervention strategy in Colombia that is triggered when the number of critical patients exceeds a threshold. They showed that this strategy could significantly reduce the total number of cases and prevent up to 97\% of deaths from COVID-19. Cancino {\it et al} \cite{cancino} proposed a model to study several measures to contain the epidemic spread in Chile and obtained that a combination of lockdown and contact tracing measures is the most effective strategy to reduce the peak demand for ICU beds.}
{In the case of Argentina}, Romero {\it et al} \cite{romero_lockdown} simulated an agent-based model for COVID-19. The authors analyzed the evolution of four different pandemic scenarios, with different levels of restriction in population mobility, and they found that social isolation is the measure that has more impact in the spread of the virus. In the same way, in \cite{neid_schools}, Neidhöfer \textit{et al} found that early school closures effectively helped in reducing the mortality rate in Argentina, Italy and South Korea. For its part, Torrente {\it et al} \cite{torrente_affec_react} showed the psychological impacts of the pandemic and {the} early measures implemented, in the form of substantial anxious and depressive symptoms. Moreover, Figar {\it et al} \cite{quiros_prevalence} conducted a seroprevalence study in one of the largest slums in CABA and found, three months after the first reported case, a prevalence of 53.4$\%$. Mobility was also analyzed in Buenos Aires City, as a proxy for effectiveness of lockdown measures. For instance, in \cite{ahumada_mobil}, Ahumada {\it et al} found a delay of 8 days between changes in mobility and reported cases, while deaths follow cases from 16 to 19 days after. 

Beyond the measures imposed by the government at a national level, each city in Argentina has the autonomy of taking different actions against the disease \cite{mdp_lock}. Crucial factors such as hospital capacity, demographic structure, and economic activities, among others, strongly vary between cities. Therefore, local authorities must adapt national guidelines to their own region and determine proper criteria in the decision-making process. For this reason, our study is focused in Mar del Plata (MDP), the main city of the Municipality of General Pueyrredón in Argentina. With this work we aim to provide a key study that will help local governments in decision-making and implementation of mitigation strategies, as well as estimating the medical resources required to face the consequences of such decisions. 

The social and demographic structure of the population is
generally reflected in heterogeneous contact patterns among
individuals, where age is one of the main determinants of these mixing
patterns. For instance, children tend to spend more time
with children and members of their household; active adults mix with
individuals in their workplace; and so forth. For this reason, a disease
transmission route may have mixing patterns among individuals of
different age \cite{ferguson2020report}. To account for this
contact structure, we use census collected data given by national agencies \cite{censo}. We divide the population into age groups and compute different contact matrices associated to different settings -household, workplace, school and general
community \cite{Mossong,fumanelli2012inferring}. 

Once we obtain the Mar del Plata {contact} structure, we consider an extension of the
compartmental Susceptible-Exposed-Infected-Recovered (SEIR) model
\cite{pastor2015} -known as SEIRD- to estimate the evolution of {the} epidemic. Here, $D$ takes into account the individuals that die as a consequence of the disease. Also, we differentiate between symptomatic and asymptomatic infected
individuals, considering different levels of illness severity for symptomatic patients. As in the real-life scenario, some individuals may not require medical attention at all (mild) and only will have to stay for a certain time at home, while others (severe) will require hospitalization \cite{meyer_hospital}.

{ In the context of the COVID-19 pandemic with such devastating consequences worldwide, the development and distribution of a new vaccine are crucial. However, reaching this ``end game'' could take too long and cost a considerable amount of deaths. For this reason, the study of non-pharmaceutical strategies to gain time is urgently needed to cease the imminent advance of the disease} \cite{vespi_strategy,ana_covid,meyer_strategy}.
{Unfortunately}, measures such as long lasting quarantines also carry serious psychological and economic consequences, that must be considered. Related to this topic, Meidan {\it et al} \cite{meidan2020} proposed a novel
alternating quarantine strategy (AQ). In this strategy the population is {separated}
into two groups, which alternate successively between
quarantine and regular activity in a bi-weekly cycle. They found that
this measure can be useful as a primary mitigation strategy, with a
comparable impact to that of a strict population-wide
quarantine. Furthermore, the weekly relief allowing people an outlet
to continue their activity for half of the time may, itself, increase
cooperation levels. {In other words}, while a complete lock-down is extremely
stressful for the individual, the alternating bi-weekly quarantine
routine relaxes the burden, and may encourage compliance. In addition, Meidan \textit{et al} \cite{meidan2020} provided detailed guidelines to apply this strategy in a real scenario.

{In this paper, we propose an age-stratified compartmental model for COVID-19 with an alternating quarantine strategy in which a fraction of the population can move freely at any time ($q_0$) while the rest is equally divided into two groups and goes through successive periods of normal activity and lockdown, each lasting $\tau$ days. We extensively explore the relationship between $q_0$ and $\tau$ and find that regardless of the value of $q_0$, the optimal period that minimizes the total number of cases is approximately $\tau=7$. On the other hand, an issue for low and middle-income countries is that a mass random testing strategy can be expensive to implement and would be a misuse of resources when the number of infected individuals is low. Therefore, we propose in our model a testing strategy that is triggered when the number of cases is above a threshold. We obtain that as the test rate $r_0$ increases, the total number of cases decreases, but this effect is diluted for high values of $r_0$. Finally, we study the combination of the testing strategy and the alternating quarantine and estimate the required number of tests, as well as the demand for ICU beds. In the worst case, our model estimates that four thousand deaths would occur, throughout the entire epidemic, of which 30\% could be avoided with proper intensive care. Nonetheless, if an aggressive massive testing is implemented, it would be possible to increase the percentage of people that can circulate freely, being the medical facilities required to deal with the additional critical patients relatively low.
Furthermore,} it is important to note that, even though we chose MDP and calibrated the model for COVID-19, our study can be adapted to any region or country where demographic information is available, and for diseases transmitted similarly by contact.

{Our work is organized as follows. In section 2, we present our differential equations that govern the spread of the disease and the equations for the alternating quarantine and massive testing strategies. In section 3, we present our results for the following strategies: i) only a massive testing strategy is conducted, ii) only a quarantine strategy is applied, and iii) both strategies are applied. Finally, in section 4, we present our conclusions.}

\section{Model}
\subsection{Compartments}

In this work, we use an extension of the SEIR model to study the
spread of {the} COVID-19 in MDP. In this model individuals can be
assigned to four different compartments, depending on their health
status. Susceptible ($S$) individuals are healthy and prone to be
infected, becoming exposed ($E$) when this occurs. At the $E$ stage,
individuals have not developed symptoms yet and they are not contagious. With the onset of symptoms, exposed individuals become
infected ($I$) and can propagate the disease. Finally, recovered ($R$)
individuals have acquired immunity and no longer propagate the
disease. One modification we make to this model is the addition of a
compartment of deceased ($D$) individuals, which are usually not
distinguished from the recovered individuals. Additionally, we include
a compartment of undetected ($U$) individuals, which takes into account the infected people who do not seek medical attention because they are asymptomatic or have very mild symptoms. These individuals are able to propagate the
disease (several publications \cite{gudbjartsson2020spread,
 mizumoto2020estimating, rothe2020transmission,
 nishiura2020estimation, kimball2020asymptomatic} have reported the
importance of asymptomatic individuals in the transmission of COVID-19). 

As we are especially interested in estimating the medical resources
that should be allocated to sick individuals, we model their evolution
by recognizing different levels of illness severity. On the one hand,
some cases develop a mild ($M$) version of the disease. Thus,
they may not require
medical care, but must remain isolated at their homes or in isolation sites prepared for this purpose, until they recover. On the other hand, some
individuals develop more serious symptoms and are hospitalized
($\mathcal{H}$) for control. Finally, we divide the hospitalized
patients into four categories, based on the fact that the evolution of
each type of patient has particular characteristic times and requires
different resources from the hospital:
\begin{enumerate}
\item $H$: those in general beds who will fully recover from the disease,
\item $H^*$: those in the intensive care units (ICU) who will fully recover from the disease,
\item $H^\dagger$: those in general beds who will die,
\item $H^{\dagger*}$: those in the ICU who will die.
\end{enumerate}
Note that we use $\mathcal{H}$ when referring to all compartments of
hospitalized individuals together, as a set. For the different
subgroups of $\mathcal{H}$, i.e, the $H$'s, we use a dagger
($\dagger$) as a notation to mark the compartments that evolve to the
compartment $D$ of deceased individuals, while an asterisk ($*$)
denotes the compartments of individuals that make use of intensive
care units (ICU).

\begin{figure}
\begin{center}
\includegraphics[width = 0.95\textwidth]{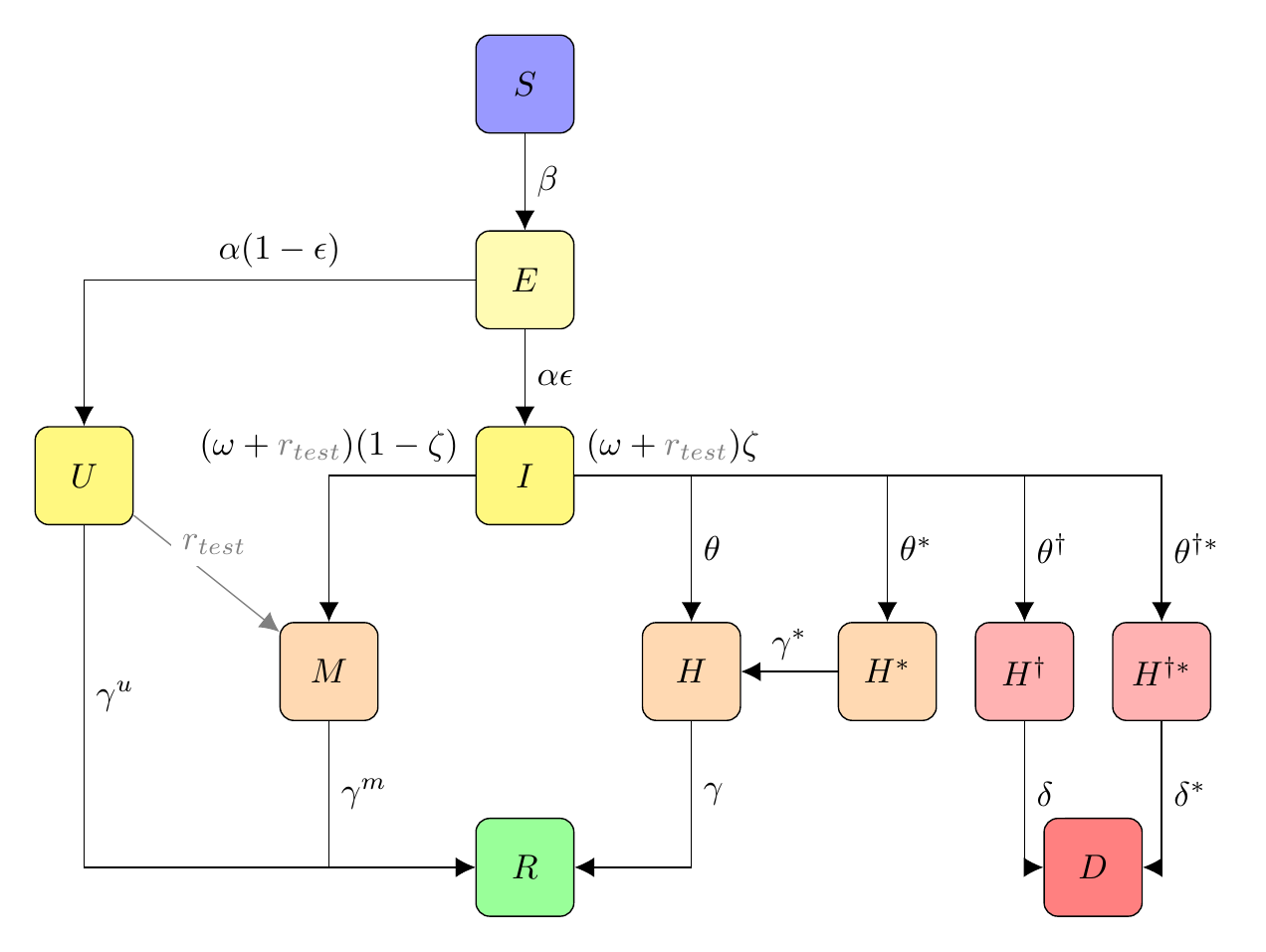}
\end{center}
\caption{{\bf Flowchart of our epidemic model.}
The susceptible population $S$ evolves to the exposed compartment $E$ with an effective rate {of contagion }dependent on $\beta$ because of the interaction with infected individuals in compartments $I$ and $U$. The exposed population becomes infected with rate $\alpha$; a fraction $\epsilon$ develops notorious symptoms and progresses to the $I$ compartment, while the remaining goes to the undetected compartment $U$. Symptomatic infected individuals pass to one of the compartments among the $M$ and $\mathcal{H}$ with a rate $\omega$. On the right side of the flow chart we have the four hospitalized compartments, which are differentiated with the presence of superscripts. Finally, individuals can end up in one of the two final states, recovered (R) or deceased (D). Note that the gray arrow from the $U$ to the $M$ compartment and the term $r_{test}$ in the outflow rates of the $I$ compartment correspond to the testing strategy explained in subsection 2.4.}
\label{flow}
\end{figure}

In Fig. \ref{flow} we show a flowchart of our model. At the
beginning, most of the population starts in the $S$ compartment. With an effective rate dependent on $\beta$, susceptible individuals evolve to the $E$ compartment, because of the interaction with the infected
population. The exposed population becomes infected with rate
$\alpha$; a fraction $\epsilon$ develops symptoms and progresses to the
$I$ compartment, while the remaining goes to $U$. We consider that symptomatic individuals do not immediately become aware of their health state, thus, we include a rate $\omega$ for the progression from $I$ to the $M$ and $\mathcal{H}$ compartments. Only a fraction $\zeta$ of $I$ individuals are hospitalized, while the remaining are isolated at home ($M$ compartment). These isolated individuals fully recover from the
disease with rate $\gamma^m$. {Note that the $M$ compartment accounts for individuals with mild symptoms who have been instructed by a healthcare professional to isolate themselves at home because they have mild symptoms. These cases are included in the total number of COVID-19 cases reported by the authorities. On the other hand, the $U$ compartment includes those undetected individuals: without symptoms or with mild symptoms who do not self-isolate, and can therefore spread the disease to others.} On the right side of the flow chart we
have the four compartments of hospitalized patients, which are
differentiated {by} the presence of superscripts. The population going
to these compartments is determined by the values of the corresponding
fractions $\theta, \theta^*, \theta^\dagger$ and
$\theta^{\dagger*}$, {where $\theta + \theta^* + \theta^\dagger + \theta^{\dagger*} = 1$}. On the other hand, the exit of these compartments
is done with different $\gamma$-rates, in case the patients recover,
or with different $\delta$-rates, for fatal cases. Note that patients
in the ICU who will recover, i.e., individuals in the $H^*$ compartment,
first flow to the $H$ compartment. 

The choice of multiple $\mathcal{H}$ compartments ensures that the
fractions of patients who fully recover from the disease and those who
die are consistent with the values reported in the literature. These
fractions are only determined by $\theta, \theta^*, \theta^\dagger$,
and $\theta^{\dagger*}$, and not by the output rates ($\gamma$- and $\delta$-rates). A summary
of the states and parameters can be found in \nameref{S1_Table} and \nameref{S2_Table}.

\subsection{System of equations}
In an effort to better represent the society, we will divide the population into age groups and incorporate contact matrices
\cite{prem2017projecting,y2018charting} to our model. These matrices indicate the mixing patterns between individuals of the population, i.e., the frequency
with which people of a certain age interact with each other, and with people from different age groups. This enables us to estimate better the
impact {of the pandemic} over the health system, and the medical resources that will be required. {For the particular case of Mar del Plata}, we use the census data collected by official organizations \cite{censo} and the inferring method developed in \cite{fumanelli2012inferring} to compute the matrices associated with different kinds of social settings: households, workplaces, schools, and general community. A summary of the inference method can be found in \nameref{S1_Calculation}.

Each age group has its own set of equations describing
the evolution of the disease spread, which is indicated by the
subscript $i$. Note that some parameters are age-dependent while
others are not, indicated by the presence or absence of this
subscript. The evolution of each group is given by:

\begin{align}
 \frac{dS_i}{dt} &= - \mathcal{F}_{i}(t) \; S_i,\\
 \frac{dE_i}{dt} &= \mathcal{F}_{i}(t) \; S_i - \alpha \; E_i,\\
 \frac{dI_i}{dt} &= \alpha \; \epsilon \; E_i - \omega \; I_i,\\
 \frac{dU_i}{dt} &= \alpha \; (1-\epsilon)\; E_i - \gamma^{u} \; U_i\\
 \frac{dM_i}{dt} &= \omega \; (1-\zeta_i) \; I_i - \gamma^{m} \; M_i,\\
 \frac{dH_i}{dt} &= \omega \; \zeta_i \; \theta_{i} \; I_i + \gamma^{*} \; H_i^{*} - \gamma \; H_i,\\
 \frac{dH_i^*}{dt} &= \omega \; \zeta_i \; \theta_{i}^{*} \; I_i - \gamma^{*} \; H_i^{*},\\
 \frac{dH_i^{\dagger}}{dt} &= \omega \; \zeta_i \; \theta_{i}^{\dagger} \; I_i - \delta \; H_i^{\dagger},\\
 \frac{dH_i^{\dagger *}}{dt} &= \omega \; \zeta_i \; \theta_{i}^{\dagger *} \; I_i - \delta^{*} \; H_i^{\dagger *},\\
 \frac{dR_i}{dt} &= \gamma^{u} \; U_i + \gamma^{m} \; M_i + \gamma \; H_i,\\
 \frac{dD_i}{dt} &=\delta \; H_i^{\dagger} + \delta^{*} \; H_i^{\dagger *},
\end{align}
where the force of infection $\mathcal{F}_{i}(t)$ is defined by
\begin{align}\label{eq.Forc1}
\mathcal{F}_{i}(t) &= \sum\limits_{j} C_{ij} \frac{\beta \; (I_{j} +
 U_{j})}{N_{j}}\ ,
\end{align}
$C_{ij}$ is the element of the contact matrix which corresponds to the interaction strength between individuals of age {groups} $i$ and $j$, and $N_j$ the total number of individuals of age {group} $j$.
{The} \nameref{S1_Table} and \nameref{S2_Table} show the parameters of our model and their values. We denote $\beta$ as $\beta_\textnormal{free}$ for the case of free propagation without any intervention. 

{
For our stochastic simulations, we model a population with $N_j$ agents in each age group (see \nameref{S2_Table}), using a discrete time approach, where time advances in fixed $\Delta t$ time steps, and the state of all agents is updated synchronously \cite{fennell2016limitations}. In this work, we use $\Delta t=0.1$ days.}

\subsection{Alternating quarantine}

{In countries where the vaccine is not yet available to all citizens,} it is of major importance that society
implements different non-medical policies to prevent and mitigate the damages that
COVID-19 may cause. Being able to estimate the amount of resources
that will be needed, as the epidemic progresses, is a key aspect that
authorities should seek in order to implement appropriate policies. In
this section, we present a particular intervention strategy that could
be helpful in accomplishing a rather controlled progression of
COVID-19 in any population, although we focus in MDP, for which we use its census data.

The alternating quarantine (AQ) strategy consists in
dividing the population into different groups that face alternate
periods of activity/isolation in terms of {their} socioeconomic
activities (education, work, recreation, transportation, etc.). There
is one group that is permanently active (labeled as $q_0$). The remaining population is separated into two equal parts ($q_1$ and $q_2$), one that is active and in contact with the $q_0$ group, and one that implements a strict lockdown (these isolated individuals can not propagate the disease). After an established period of time $\tau$, $q_1$ and $q_2$ switch places. The parameter $\tau$ will be tuned in order to find the best possible scenario for diminishing the spread of the disease.

For this strategy, we subdivide each compartment $S_i$, $E_i$, $I_i$ and $U_i$ into three groups $q_0$, $q_1$, and $q_2$ (the labels represent also the fraction of the population in each group, so that $q_1=q_2=\frac{1-q_0}{2}$). Thus, adapting Eq.~(\ref{eq.Forc1}) to the case of alternating quarantine, the force of infection for each group must be rewritten as follows:

\begin{align}
\mathcal{F}_{i}^{\ \mathcal{Q}}(t) &= \sum\limits_{j} C_{ij} \frac{\beta^{\mathcal{Q}} \;
 (I_{j}^{\ \mathcal{Q}} + I_{j}^{\ q_0} + U_{j}^{\ \mathcal{Q}} +
 U_{j}^{\ q_0})}{N_{j}}\ ,\\
\mathcal{F}_{i}^{\ q_0}(t) &=
\mathcal{F}_{i}^{\ q_1}(t) +\mathcal{F}_{i}^{\ q_2}(t),
\end{align}
where $\mathcal{Q} = \{q_1, q_2\}$. The rates of infection are given by:

\begin{align}
 \beta^{\ q_1} &= 
 \beta_\textnormal{free}\ \Theta\left[\sin\left(\frac{\pi t}{\tau}\right)\right]\ ,\\
 \beta^{\ q_2} &= 
 \beta_\textnormal{free}\left(1-\Theta\left[\sin\left(\frac{\pi t}{\tau}\right)\right]\right)\ ,
\end{align}
 where $\Theta(\cdot) \in [0,1]$ is the Heaviside function {(with $\Theta(0) = 1$)}, so that $\Theta\left[\sin\left(\frac{\pi t}{\tau}\right)\right]$ is a periodic square function of period $2\tau$. Note that $\beta^{q_1}$ and $\beta^{q_2}$ are perfectly out of phase.

\subsection{Population-wide testing}

{In the context of an epidemic outbreak, identifying infected individuals is crucial to attenuate the imminent propagation as it allows authorities to isolate specific carriers of the disease \cite{eckerle_cont_trace}.} Particularly, the detection
of infected individuals with very mild symptoms or no symptoms at all is a difficult
task to achieve, since these individuals do not know they are ill
and, therefore, do not seek medical attention. It is extremely
important to find and isolate them as quickly as possible. For this
reason, along with the AQ strategy, we implement a random massive testing strategy with rate $r_{test}$ in which individuals in the $U$ compartment flows to the $M$ compartment. Similarly, symptomatic infected individuals $I$ will be detected by this strategy and admitted to the hospital or isolated, as previously explained. Mathematically, this means replacing in Eqs. (3), (5)-(9) $\omega$ by $\omega + r_{test}$.

{A disadvantage of a massive testing strategy is that when the number of active cases in the community is low, most tests are negative and cannot be reused. Therefore, in our testing strategy, we consider that the tests are conducted only when the total number of infected individuals at time $t$, $I(t)$, exceeds a threshold $I_0$. More specifically, we propose the following test rate:
\begin{eqnarray}\label{eq.rtestThet}
r_{test}(t)=r_0\;\Theta(I(t)-I_0).
\end{eqnarray}
According to this strategy, if the number of infected individuals exceeds $I_0$, the population is tested at a rate $r_{test}=r_0$, reducing transmission of the virus. On the other hand, when the number of infected individuals is below $I_0$, the testing strategy is suspended ($r_{test}=0$). In particular, when $I_0\to 0$, tests are always conducted at rate $r_0$, whereas in the limit $I_0\to \infty$, tests are never conducted.}

\subsection{ICU capacity}

In general, local and national governments make public health
decisions based on the projected availability of medical resources, such as: ICU beds, {general beds,} ventilators, and the specialized workers {needed to} operate the equipment and treat the patients \cite{phillips_alloc,meyer_distancing}. The exponential
growth of COVID-19 spread may cause a {quick} depletion of
resources and the saturation of the health system. {In consequence, an important number of avoidable deaths may occur.} {Therefore, it is relevant to predict the evolution of ICU beds in hospitals and to foresee the consequences of their depletion.}
To include the impact of ICU saturation in our deterministic equations, first, we numerically solve Eqs. (7), (9), and (11) for each age group $i$. More specifically, we use the forward Euler method with a time step of $\Delta t=0.1$ [days] to estimate the number of individuals in compartments $H_i^*$ and $H_i^{\dag *}$ at time $t+\Delta t$. Before we update the new values of $H_i^*$ and $H_i^{\dag *}$, we verify for each age group that the number of ICU beds will not be exceeded. If this condition is not satisfied, the excess demand for ICU beds is diverted to the $D_i$ compartment while the computed values of $H_i^*$ and $H_i^{\dag *}$ at time $t+\Delta t$ are reduced, taking into account their proportional contribution to the excess demand.

{In the particular case of MDP,} despite the fact that the local authorities have not reported an official number of available ICU beds \cite{quedig_pedido}, we estimate that there are roughly 180 ICU beds in the municipality, and approximately 60 of these beds are available to accommodate COVID-19 patients \cite{pag12, quedig_montenegro}.

\section{Results and discussion}

As mentioned in the Introduction, on March 3, 2020, Argentina's Ministry of Health confirmed the first case of COVID-19 in the country and two weeks later a strict national lockdown was imposed, which was gradually relaxed in the following months. Although this measure was crucial to avoid an overwhelmed healthcare system and to give hospitals time to increase their capacity, it also had a negative impact on the economy. In this section, we study the scenario in which a massive testing and AQ strategies are applied on the population since September 30, 2020, dividing the analysis into four subsections. In the first one, we estimate: i) the value of $\beta_\textnormal{free}$ and ii) the initial values of individuals in each compartment on September 30, 2020, by fitting the actual death curve in the city of Mar del Plata. In the second and third subsections, we analyze the AQ and massive testing strategies separately from September 30 onwards, and in the last subsection we discuss their joint effect on the epidemic spreading. Except for the first subsection, we always compute the relevant magnitudes for 300 days, i.e., until June 7, 2021, because the number of infected individuals becomes negligible after this period, for the parameter values explored in this work\text{*}. 
\let\thefootnote\relax\footnotetext{\text{*}: The code for our simulations is publicly available on Github \cite{github}.}

\subsection{{Estimation of $\beta_\textnormal{free}$ and the initial conditions}}

\begin{figure}
\begin{center}
\includegraphics[width = 0.6\textwidth]{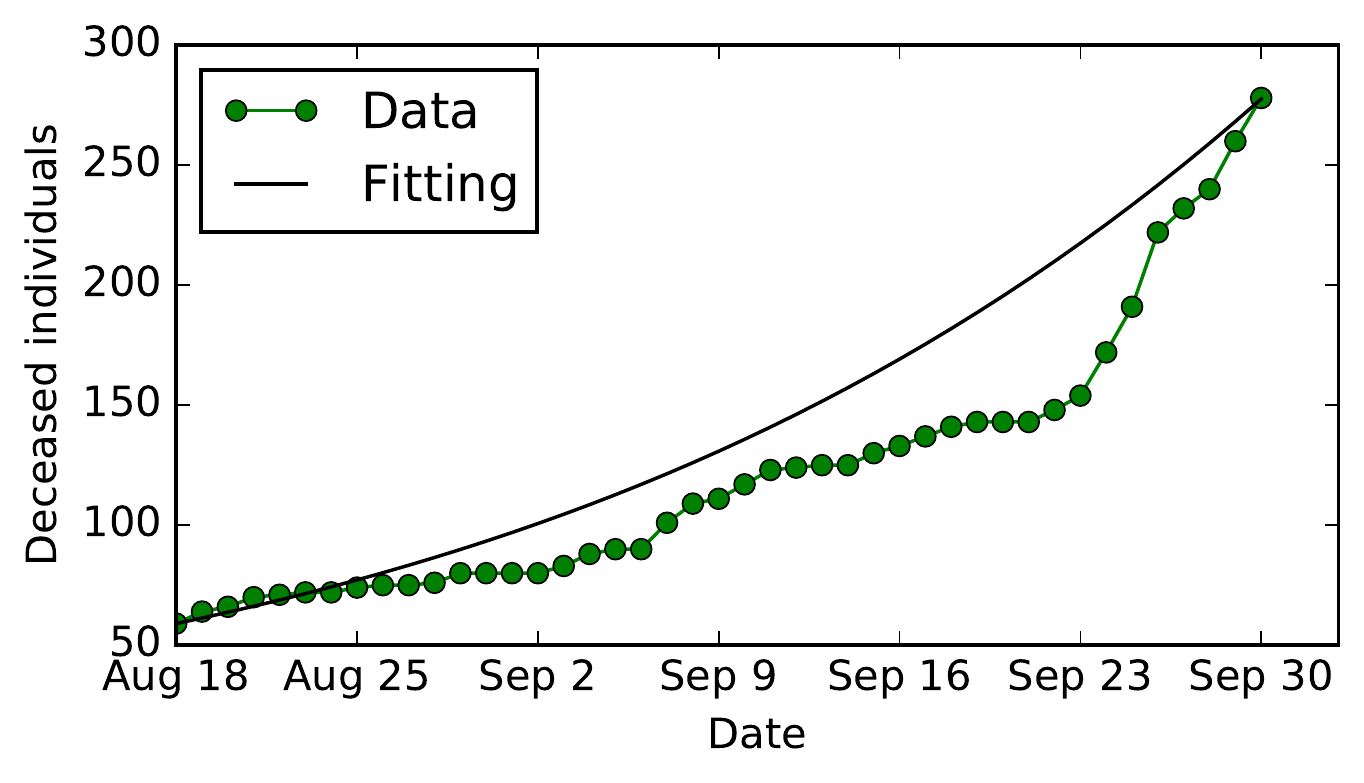}
\end{center}
\caption{{\textbf{Fitting of the actual death curve.} The solid black line represents the theoretical curve that minimizes the mean square error with the real data (green dots). The curve corresponds to $\beta= 3.45$, which allows us to estimate the initial conditions of the rest of the compartments.\label{deathfit}}}
\end{figure}

{In contrast to other countries where the disease spread freely for several weeks with a non-negligible number of daily cases, Argentina rapidly imposed measures two weeks after the first case was reported. Consequently, there are insufficient data to estimate the infection rate $\beta_\textnormal{free}$ for free propagation in Argentina, and particularly in Mar del Plata. Although the goal of our paper is not to model the beginning of the disease spreading but from September 30 onwards, the parameter $\beta_\textnormal{free}$ is relevant in our work since the infection force is proportional to $\beta_\textnormal{free}$, for the non-quarantined individuals in the AQ strategy (see Eqs.(13)-(16)). As the basic reproduction number at the beginning of the propagation in several countries was reported to be approximately $R_0=2.5$ \cite{wu2020nowcasting, li2020early, riou2020pattern}, we fit $\beta_\textnormal{free}$ for our model by using the next generation matrix method \cite{diekmann2010construction} such that the value of $R_0$ for Mar del Plata is also $R_0=2.5$ in a stage of free propagation. By doing so, we obtain that $\beta_\textnormal{free}=5.6$.

On the other hand, as of September 30, 2020, a total number of 12862 cases and 278 deaths in Mar del Plata were reported. However, due to asymptomatic cases and a low testing rate, it is to be expected that the former number was significantly underestimated. Therefore, to estimate the number of individuals in each compartment on September 30, we fit our model to the actual death curve from August 18 to September 30, obtaining $\beta=3.45$. Note that the death curve has also been used in previous studies to estimate unknown parameters \cite{mills2004transmissibility}. More specifically, to fit $\beta$, we integrate Eqs. (1)-(11) in this period for different values of $\beta$ in the interval [2.3, 8.0]. Then we choose the optimum value that minimizes the square distance between the predicted numbers of accumulated deaths and their reported values before September 30, as shown in Fig. \ref{deathfit}. After fitting our model, we compute the number of individuals in each compartment on that date.}
 
{Briefly, we obtained that on September 30, 2020: approximately 7,300 individuals were exposed individuals ($E$), 4,144 were symptomatic non-isolated individuals ($I$), 268 were hospitalized (53 in ICU beds), 6500 patients were isolated at home ($M$), and 12,000 remained undetected ($U$). In \nameref{S3_Table}, we show the estimated number of individuals in each compartment segregated by age group.}

\subsection{{Mass testing strategy without alternating quarantine}}

\begin{figure}
    \centering
    \begin{subfigure}[b]{0.475\textwidth}
        \centering 
        \begin{overpic}[width = \textwidth]{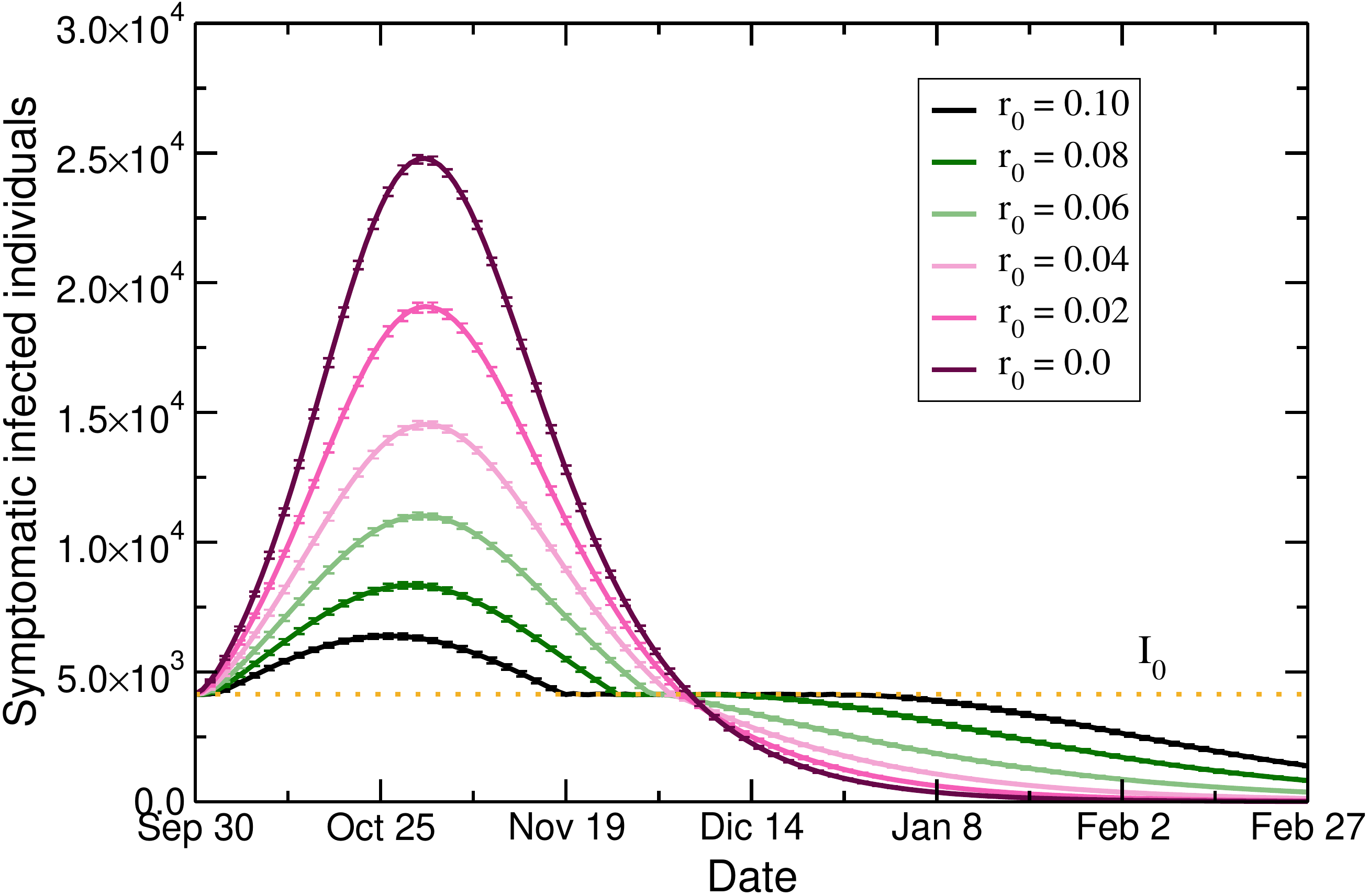}
            \put(17,56){\sffamily \big (A)}
        \end{overpic}
    \end{subfigure}
    \begin{subfigure}[b]{0.475\textwidth}  
        \centering 
        \begin{overpic}[width = \textwidth]{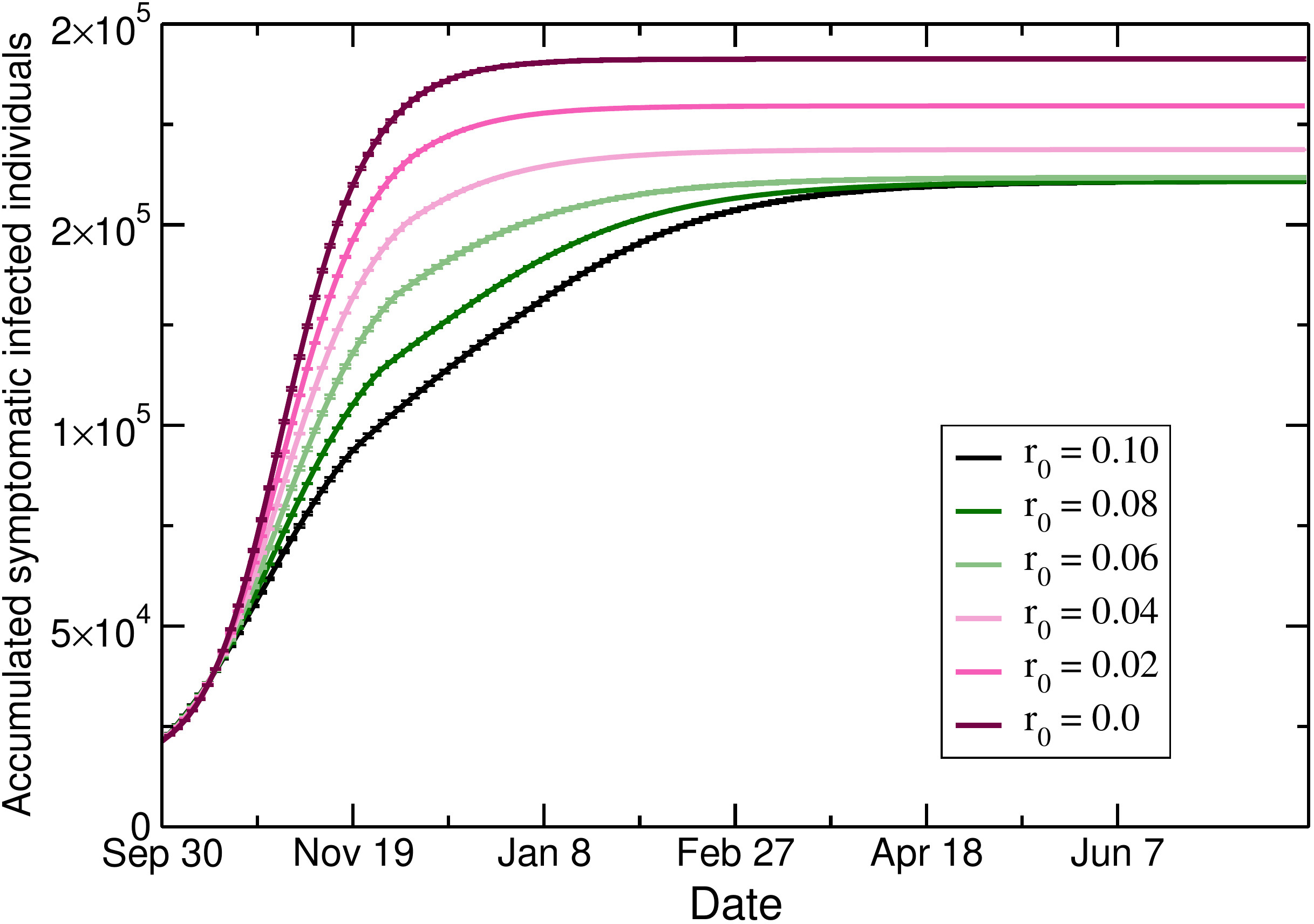}
            \put(17,56){\sffamily \big (B)}
        \end{overpic}
    \end{subfigure}
\caption{\textbf{Mass testing effect on infected population.} (A) The number of symptomatic infected individuals ($I$) and (B) the accumulated number of symptomatic infected individuals ($I_{ac}$) predicted by our model for different values of the testing parameter $r_0$ and $q_0 = 1$ (no quarantine) and $I_0=4144$. The error bars were estimated from 50 stochastic simulations.
    \label{massResult}}
\end{figure}

{In this subsection, we will analyze the effect of a massive testing strategy on the spread of the virus without implementing the quarantine strategy, i.e., $q_0 = 1$. Figure \ref{massResult}A shows the time evolution of the number of infected individuals for different values of $r_{0}$ and with $I_0=4144$ (see Eq. (17)) which is the estimated number of infected individuals on September 30, 2020 (see Sec. 3.1). It can be seen that the strategy decreases the height of the peak, and the curve flattens around $I=I_0$. On the other hand, for the time evolution of the cumulative number of cases (Figure \ref{massResult}B), $I_{ac}$, as the value of $r_0$ increases, the final number of accumulated cases $I_{ac}(t=\infty)$ decreases. This result is to be expected because patients who test positive are immediately isolated, which mitigates the spread of COVID-19 in the population. However, the effect of $r_0$ on $I_{ac}(t=\infty)$ becomes less noticeable as $r_0$ increases. To explain this behavior, in \nameref{S2_Calculation}, using a SIR model with a testing rate given by Eq.~(\ref{eq.rtestThet}), we study in detail the limit $r_0=\infty$. In particular, we show in \nameref{S2_Calculation} that as $r_0$ increases in Eq.~(\ref{eq.rtestThet}), the curves of infected individuals $I(t)$ approach the case of $r_0=\infty$, in which $I(t)$ remains constant at $I=I_0$ for a period of time, as a consequence of the activation-deactivation of the strategy around this value. During this plateau, the number of susceptible individuals decreases until the population reaches herd immunity, and from this moment on, the number of infected individuals declines exponentially. Therefore, a testing strategy with a test rate given by Eq.~(\ref{eq.rtestThet}) is effective in flattening the curve of infections and avoiding an overwhelmed healthcare system. However, if it is not combined with other measures to lower the probability of infection, then a large number of individuals will have been infected by the end of the epidemic.

In the supplementary material, \nameref{S4_Table} and \nameref{S5_Table}, we show the impact of different thresholds $I_0$ on the total number cases and tests used until June 7, 2021. Since in Mar del Plata the epidemic was under control on September 30 when the number of infected individuals was 4144, in the following sections we explore the impact of testing strategy on the epidemic only for $I_0=4144$.}

\subsection{{Alternating quarantine without mass testing}}

\begin{figure}
\centering
    \begin{subfigure}[b]{0.475\textwidth}
        \centering 
        \begin{overpic}[width = \textwidth]{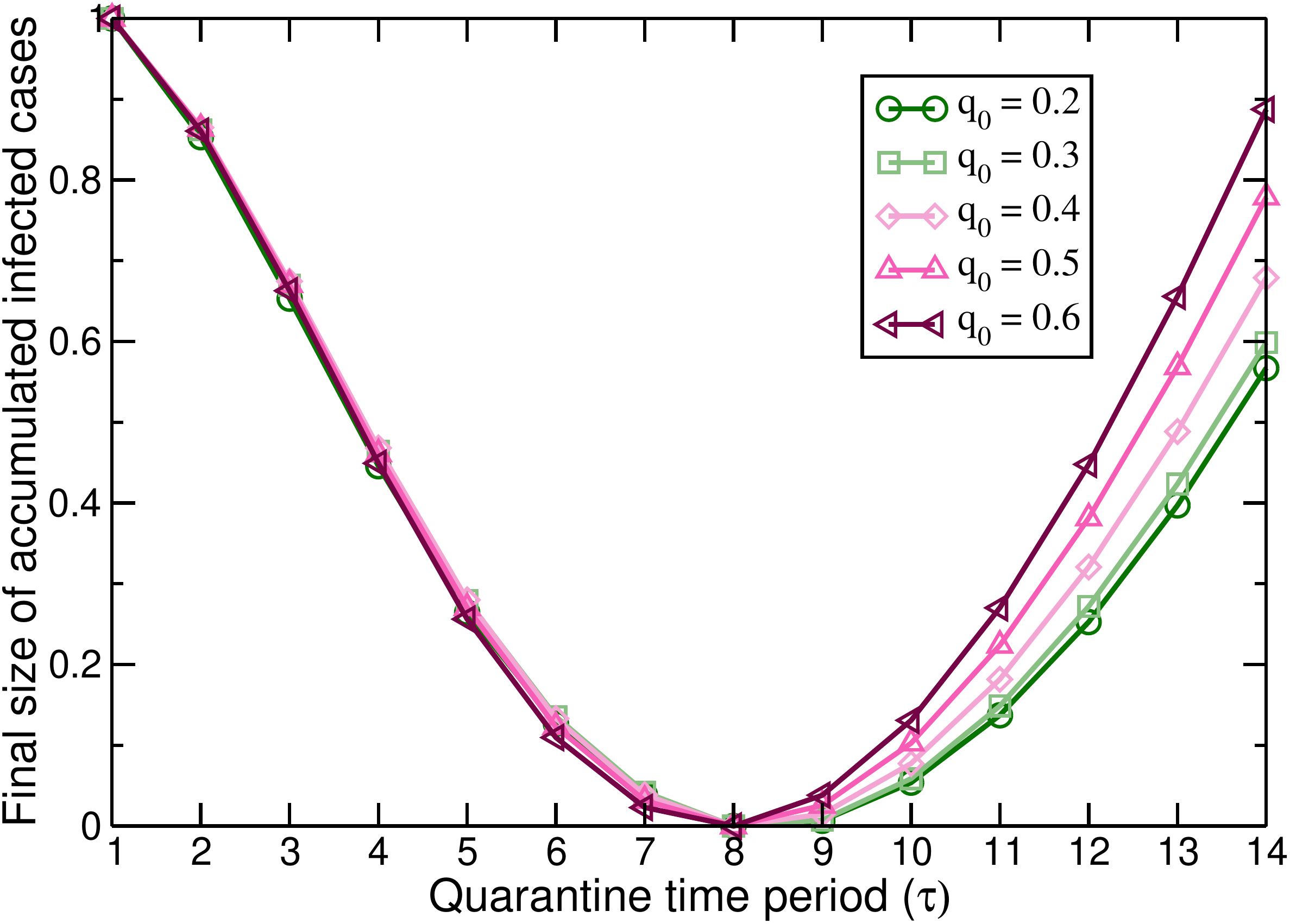}
            \put(15,15){\sffamily \big (A)}
        \end{overpic}
    \end{subfigure}
    \begin{subfigure}[b]{0.475\textwidth} 
        \centering 
        \begin{overpic}[width = \textwidth]{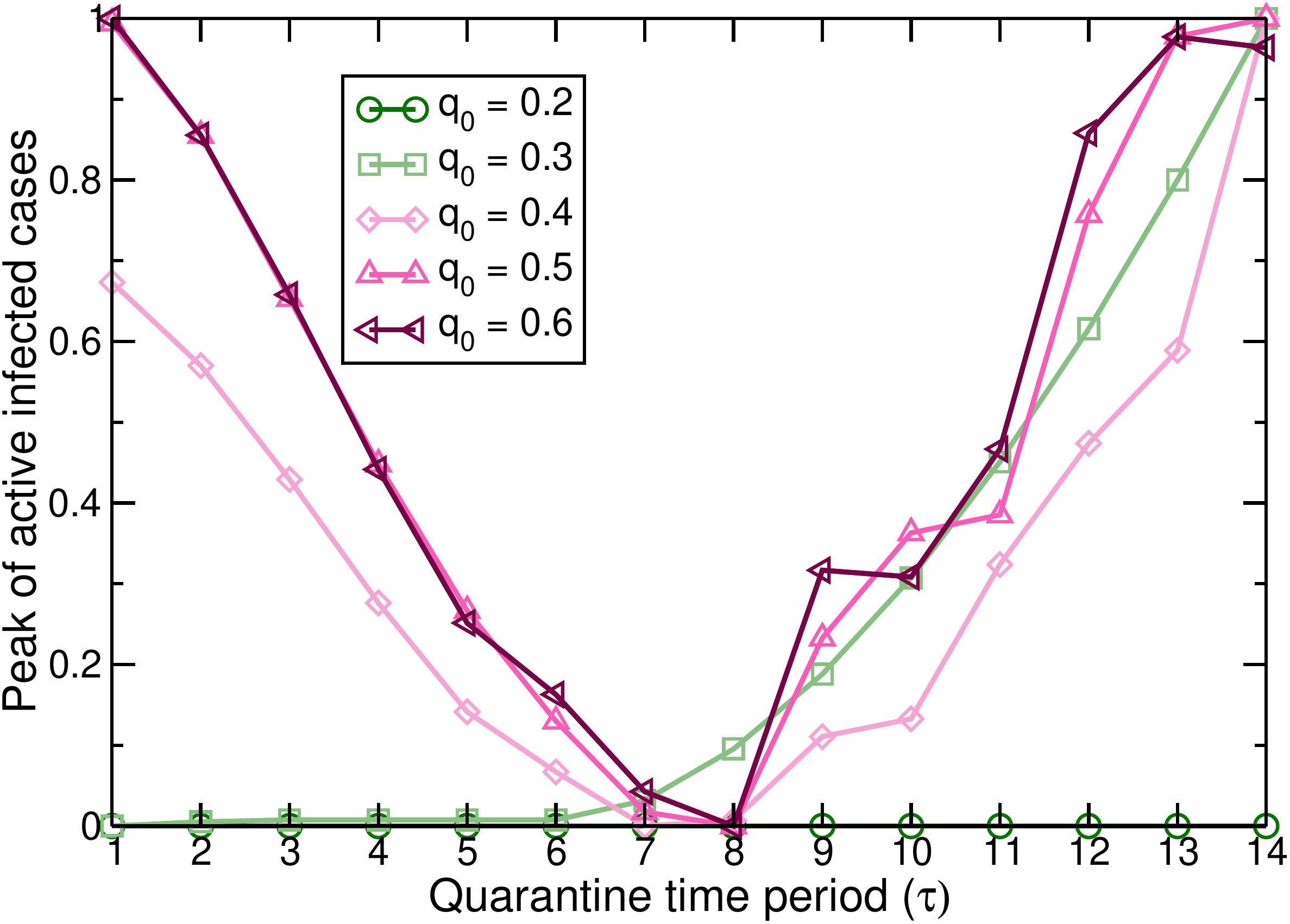}
            \put(15,15){\sffamily \big (B)}
        \end{overpic}
    \end{subfigure}
\vspace*{5mm}
\caption{{\bf Optimal value of $\tau$.}
(A) The accumulated number of infected individuals at the final state and (B) the height of the peak of active cases as a function of $\tau$ for different values of the parameter $q_0$. The values were obtained by solving numerically Eqs. (1)-(11).}
\label{tau}
\end{figure}

{Now we} study the effect of the AQ on the epidemic spreading for different values of $\tau$ and $q_0$, without implementing the testing strategy ($r_{test}(t) = 0$). To this end, we solve numerically Eqs. (1)-(11) and compute the cumulative number of infected individuals at the final state (when $I \approx 0$), and the height of the peak of active cases under the AQ routine. In Fig. \ref{tau}, these two quantities are plotted as a function of the period $\tau$, which ranges from 1 to 14 days. To easily compare the curves of different values of $q_0$, we applied a min-max normalization \cite{maxmin} to show the y-coordinates in the interval $[0,1]$. As can be seen, the optimal value of $\tau$ is weakly dependent of the parameter $q_0$, with approximately 8 days being the optimum for the accumulated cases (Fig. \ref{tau}A) and between 7 and 8 days for the peak of active cases (Fig. \ref{tau}B). From here on, we use $\tau = 7$ in our simulations as it is close to optimal, and in this way, the AQ can be in phase with the standard week.

\subsection{{Joint effect of the testing and AQ strategies}}

\begin{figure}
    \centering
    \begin{subfigure}[b]{0.475\textwidth}
        \centering 
        \begin{overpic}[width = \textwidth]{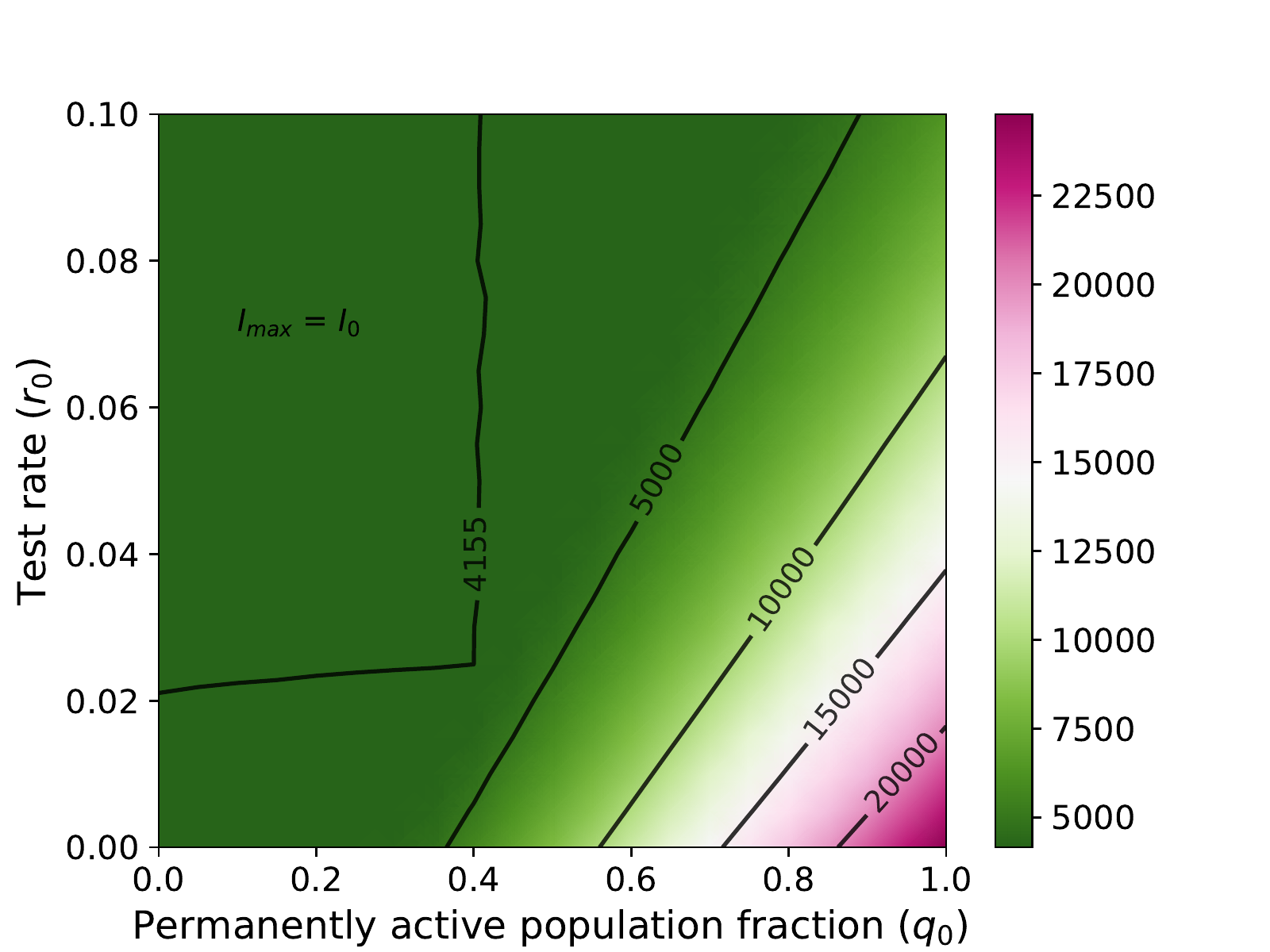}
            \put(64,55){\sffamily \big (A)}
        \end{overpic}
    \end{subfigure}
    \begin{subfigure}[b]{0.475\textwidth}  
        \centering 
        \begin{overpic}[width = \textwidth]{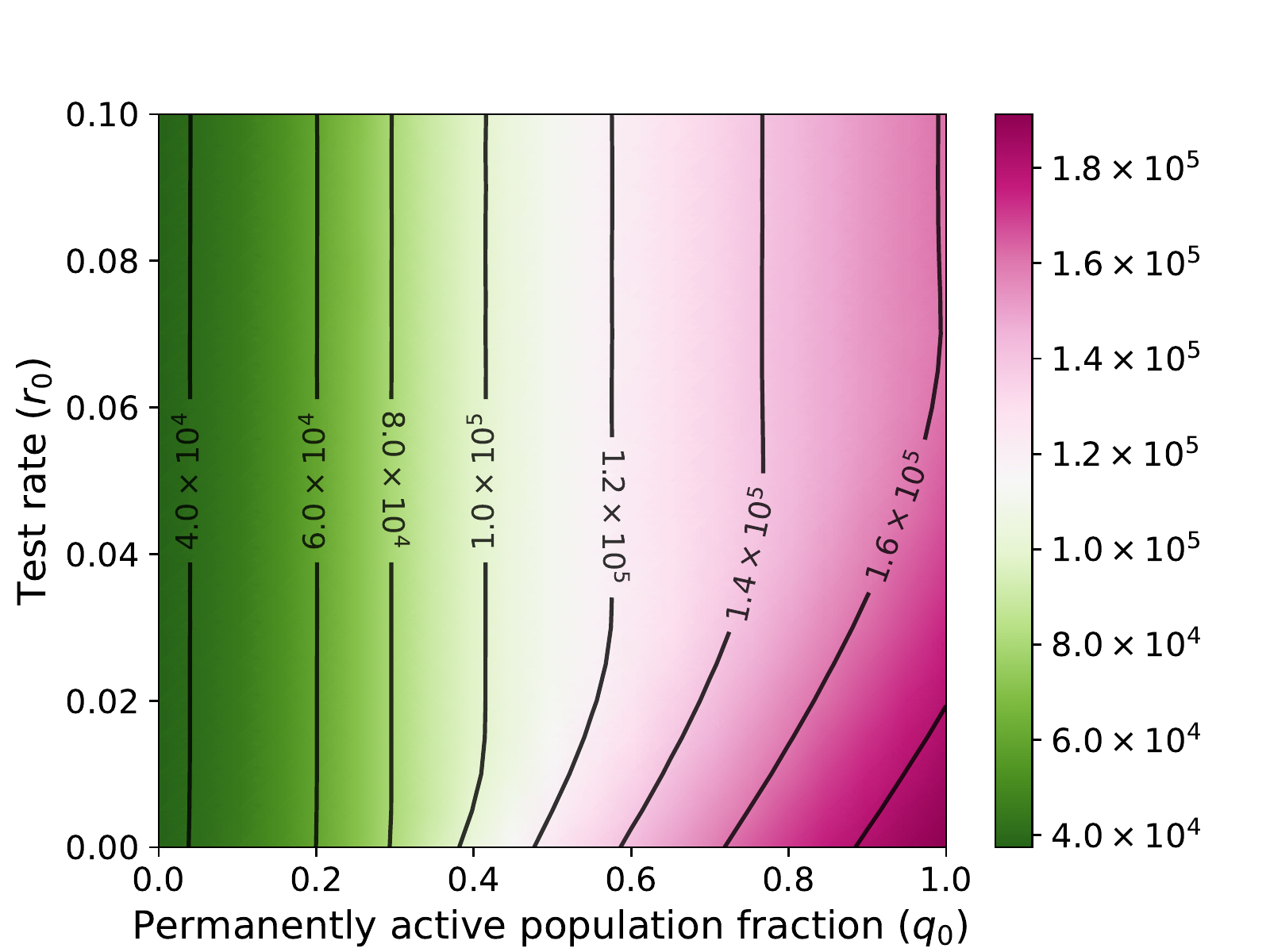}
            \put(64,55){\sffamily \big (B)}
        \end{overpic}
    \end{subfigure}
\vspace*{5mm}
\caption{{\bf Evolution of infected individuals.}
In (A) we show a heat map of the height of the peak of symptomatic infected individuals $I_{max}$ in the plane $q_0$-$r_{0}$. The black lines represent isolines. {In panel (B), in a similar way, the final sizes of accumulated symptomatic infected individuals are shown, after 300 days of simulation.}}
\label{evinf}
\end{figure}

{In this last subsection}, we will study the joint effect of the testing and AQ strategies on the epidemic spreading and how they reduce the pressure on the health system. In Fig. \ref{evinf}A, we show a heat map of the peak value of symptomatic infected individuals ($I$) as a function of the fraction of the permanently active population $q_0$ and the test rate $r_{0}$, for $I_0=4144$ (see Eq. (17) and Sec. 3.2). It can be seen that there is a region where the peak coincides with the initial value of infected individuals, implying that the curve of infected cases {can no longer grow} after applying both strategies. In contrast, for higher values of $q_0$ and lower values of $r_{0}$, the peak of infected individuals is up to {6} times higher than the threshold $I_0$. 
{On the other hand, in Fig. \ref{evinf}B, the final sizes of accumulated symptomatic infected individuals on June 7, 2021, are shown. As in the first sub-figure, the AQ strategy reduces significantly this magnitude when the value of $q_0$ decreases. It can also be noted from the vertical iso-height lines, that high values of $r_0$ cannot reduce the final size of the epidemic, which is consistent to the results shown in Section 3.2. Note that the magnitudes shown in Figs 5A-B, correspond only to detected individuals, i.e., the cases that passed through compartments $M$ and $\mathcal{H}$, but if we also include the individuals who passed through compartment $U$, we obtain qualitatively similar results. In the supplementary material we present additional results for the number of infected individuals compared with the real data after September 30 (see Fig. \ref{data}).}

\begin{figure}
 \centering
    \begin{subfigure}[b]{0.475\textwidth}
        \centering
        \begin{overpic}[width = \textwidth]{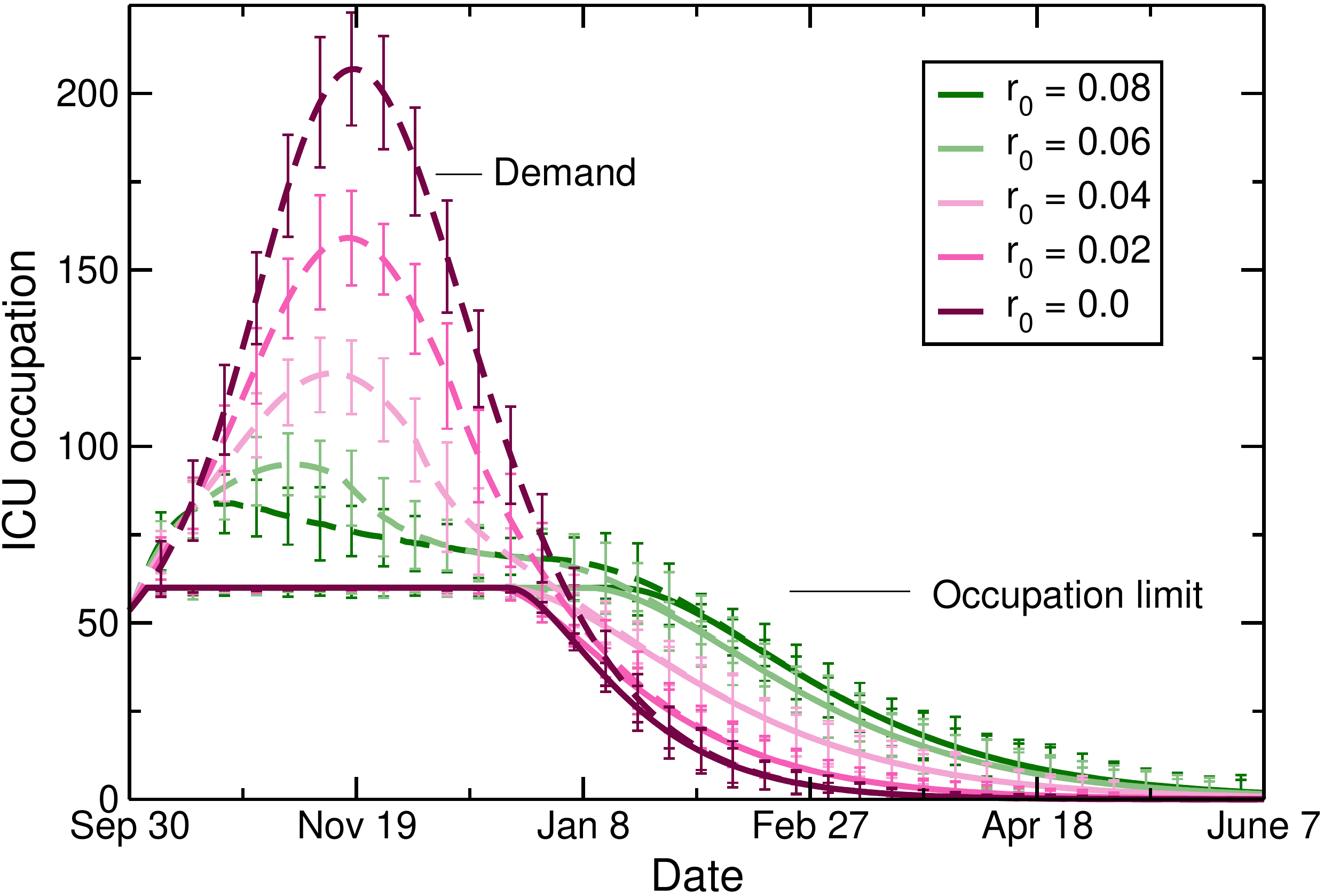}
            \put(13,15){\sffamily \big (A)}
        \end{overpic}
    \end{subfigure}
    \hfill
    \begin{subfigure}[b]{0.475\textwidth}  
        \centering 
        \begin{overpic}[width = \textwidth]{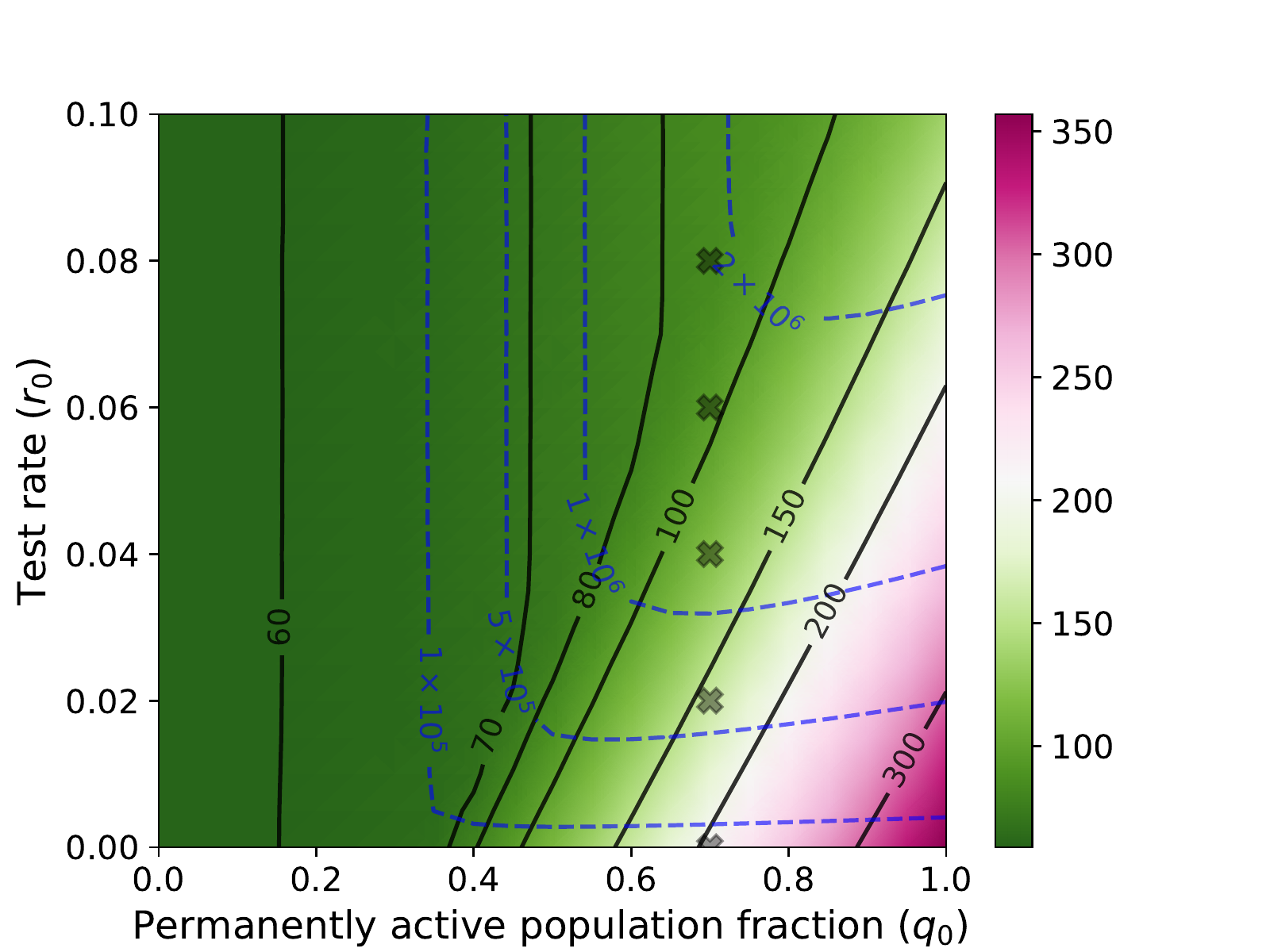}
            \put(13,15){\sffamily \big (B)}
        \end{overpic}
    \end{subfigure}
    \vskip\baselineskip
    \begin{subfigure}[b]{0.475\textwidth}   
        \centering 
        \begin{overpic}[width = \textwidth]{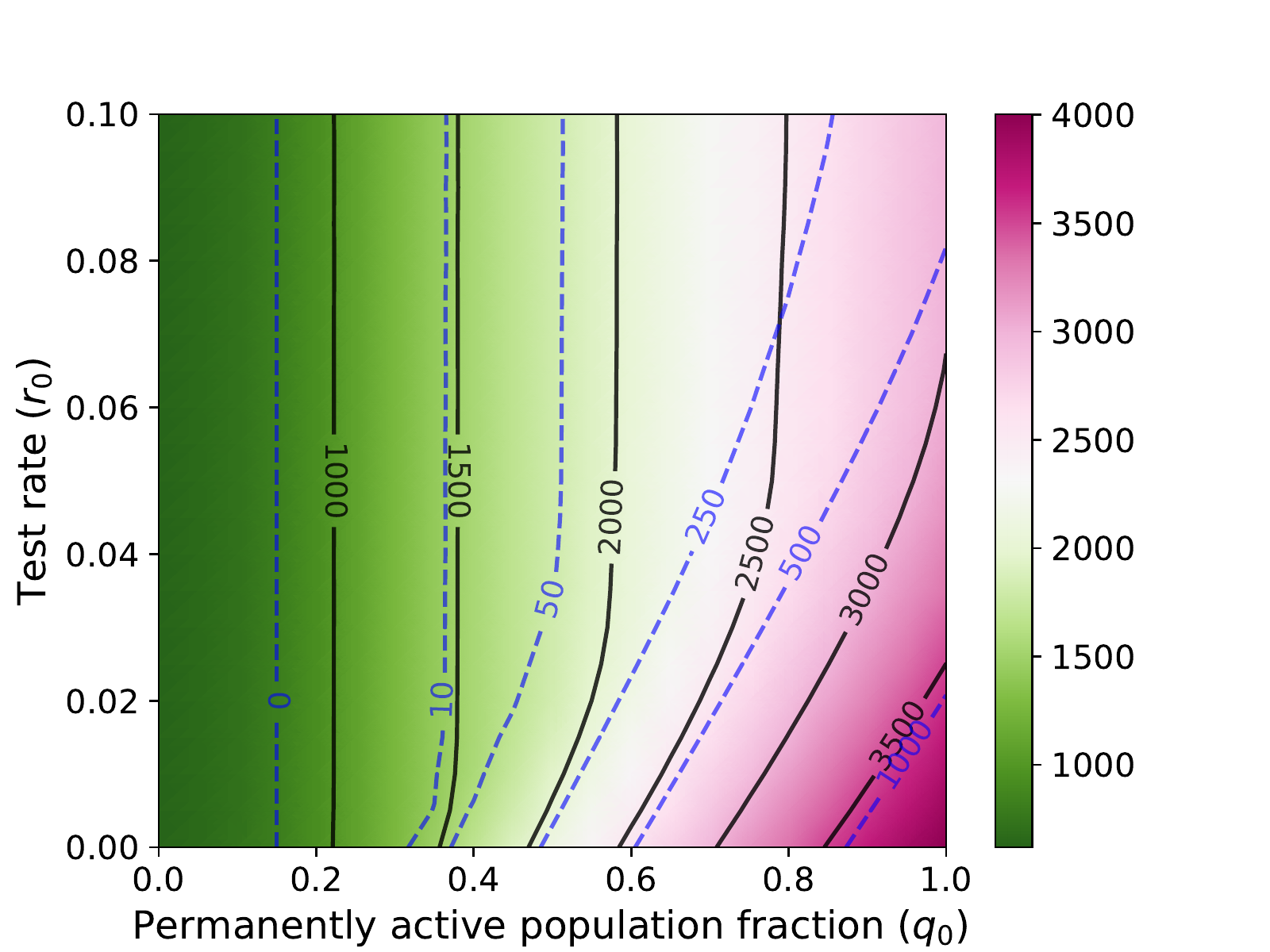}
            \put(13,15){\sffamily \big (C)}
        \end{overpic}
    \end{subfigure}
\vspace*{5mm}
\caption{{\bf Impact on the health system.}
(A) Time evolution of the ICU occupation when the number of ICU beds is 60 (solid lines) and infinite (dashed lines). Each curve corresponds to a different value of the test rate $r_{0}$, with $q_0 = 0.7$ and $I_0=4700$. The error bars were estimated from 40 stochastic simulations. (B) Heat map showing the minimum number of ICU beds ($B_{min}$) to prevent hospital saturation in the plane $q_0$-$r_{0}$. The black lines represent isolines while the cross marks serve as a reference for panel (A). The dashed blue lines correspond to the $T_{test}$-isolines (see explanation in the main text). (C) Heat map depicting the total number of deceased individuals ($D_{tot}$) at the final state of the epidemic in the plane $q_0$-$r_{0}$ when the total number of ICU beds is 60. The solid black isolines correspond to the total number of deaths ($D_{tot}$) while the dashed blue isolines represent only those deaths due to hospital saturation.}
\label{health}
\end{figure}

To further elucidate the {impact} of the epidemic spreading on the health system, we will compute the minimum number of ICU beds ($B_{min}$) to prevent an overwhelmed healthcare system. In order to estimate this magnitude we compute the time evolution of the number of critical patients when there is no restriction on the number of ICU beds, as shown in Fig. \ref{health}A, for {an increasing value of $r_{0}$ and $q_0 = 0.7$}. Note that the height of the peak of this curve indicates the value of $B_{min}$ that prevents ICU saturation. In Fig. \ref{health}B we plot a heat map of $B_{min}$ and we obtain, as expected, that the two proposed strategies reduce the pressure on the health system, i.e., $B_{min}$ decreases for larger values of $r_{0}$ and smaller values of $q_0$. Note that the region to the left side of each isoline of $B_{min}$ (solid black lines) corresponds to the values of $q_0$ and $r_{0}$ where the demand for ICU resources does not exceed the supply $B_{min}$. In the supplementary material (see Fig. \ref{overw}), we include a heat map of the period of time in which the health system is overwhelmed when the number of ICU beds is 60. 

In addition, it is also relevant to estimate the total number of tests $T_{test}$ that the municipality must purchase for a given value of $r_{0}$ and $q_0$ because this magnitude can be used to assess the budgetary impact of a massive testing strategy. To estimate this magnitude, we accumulate from our equations, the number of tests used per unit time for 300 days since September 30. The results of this procedure are shown in Fig. \ref{health}B as $T_{test}$-isolines in the plane $q_0$-$r_{0}$ (dashed blue lines). The intersection between the left region of a $T_{test}$-isoline and a $B_{min}$-isoline indicates the values of $q_0$ and $r_{0}$ that meet the budget and ICU capacity. Notably, for a given value of $q_0$, the total number of tests saturates as the testing rate $r_{0}$ increases, which is observed in the figure as a vertical $T_{0}$-isoline. 

On the other hand, since the ICU bed capacity is a critical bottleneck for the health care system, policymakers must know the total number of deaths $D_{tot}$ if the ICU capacity is not increased, and also how many deaths would be avoided if the hospitals ensured a minimum number of ICU beds to prevent saturation. In Fig. \ref{health}C we show a heat map of $D_{tot}$ at the final state when the number of ICU beds is 60. In the range of explored values, we observe that decreasing the fraction of permanently active individuals ($q_0$) reduces $D_{tot}$ more significantly than a higher testing rate $r_{0}$. Furthermore, the massive testing strategy has a meaningful effect on $D_{tot}$ only when $q_0\gtrsim0.7$. We also include in the same figure the isolines of the number of deaths avoided (dashed blue lines) in the scenario where the authorities extend the ICU capacity to prevent an overwhelmed health care system. It can be seen that for the case of a weak AQ strategy (high values of $q_0$) the total number of deaths due to the lack of ICU beds ranges from 250 to 1000 people which corresponds to up to 30\% of the total number of deaths.

\section{Conclusion}
{In this paper, we studied a combination of mitigation strategies to prevent the spread of the SARS-CoV-2 without a total restriction of activities and we applied our model to the particular case of the city of Mar del Plata.}
{This model is} based on an extended SEIR model, where detected cases are confined and do not propagate the disease. {One of the strategies} consists {of} an alternating quarantine where a portion of the population remains always active, while the rest is divided in two groups and enters a cycle of activity and isolation with period $2\tau$. We found that a time-window of $\tau = 7$ days is the best alternative as it reduces the peak of active cases and {the final size of} cumulative cases of detected symptomatic patients (a fraction of which eventually die), while it fits with the usual weekly cycle, being easy to implement. However, it should be noted that this optimal value of $\tau$ depends on the characteristic times associated with the infectious process of COVID-19, so it should be readjusted if this quarantine strategy is to be applied to mitigate other diseases. On the other hand, we studied a testing strategy on the general population with a threshold. We focused on the pressure {applied to} the health system and the death toll that an overwhelmed system may cause. We found that, given the situation in Mar del Plata as of September 30, extremely low levels of normal circulation would be necessary in order to avoid the saturation of the health system, regardless of the testing rate. The number of ICU beds necessary to prevent saturation increases with the fraction of the population that remains always active, but this number can be shortened by means of an aggressive testing strategy.
In the worst case scenario, we estimate that thousands of people may die, but would not if properly assisted. Therefore, we consider that implementing an alternating quarantine strategy with high testing rates and medium-to-low levels of free circulation is fundamental in order to change the trend of COVID-19 spread in Mar del Plata. Authorities might have to balance testing and equipment costs with mobility levels of the population, in order to arrive at a consistent strategy that people {are} able to follow, thus, ensuring compliance and reducing the economic, social and psychological effects that long and extended - or the lack of - quarantine produce{s}.

{Finally, we remark that the results shown in this work should be taken with caution because our model does not consider: viral mutations, additional measures (such as the extended use of face masks or contact tracing), imported cases, seasonality effects, or a dynamic value of $q_0$ due to quarantine fatigue. However, we believe that our model could serve as a support for future work on the effectiveness of an alternating quarantine and massive testing strategies in more realistic scenarios.
}

\clearpage
\section*{Supporting material}
\singlespacing

\renewcommand\thefigure{S\arabic{figure}}    
\setcounter{figure}{0}


\begin{figure}[htb!]
\centering
    \begin{subfigure}[b]{0.475\textwidth}
        \centering
        \begin{overpic}[width = \textwidth]{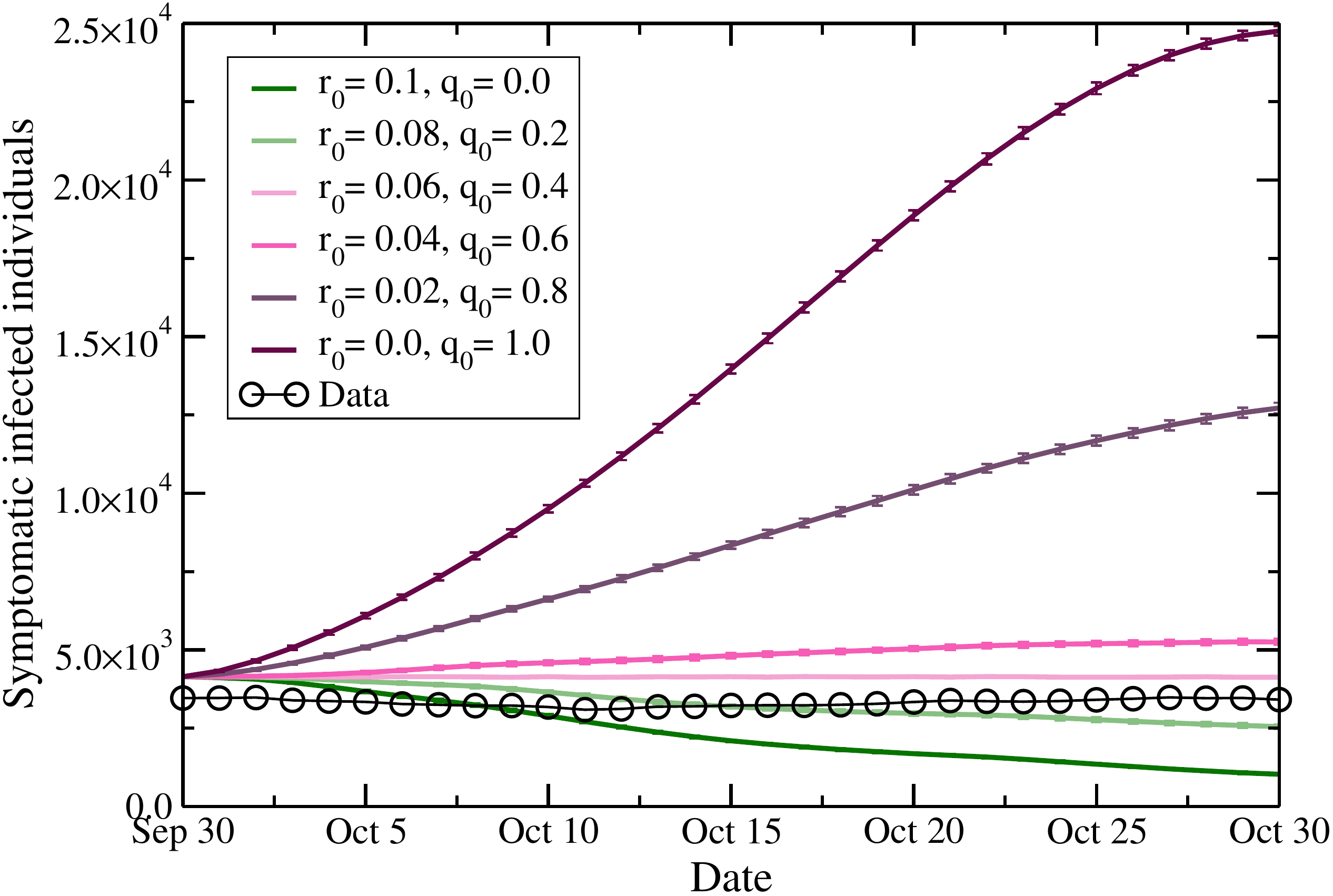}
            \put(75,52){\sffamily \big (A)}
        \end{overpic}
    \end{subfigure}
    \begin{subfigure}[b]{0.475\textwidth}  
        \centering 
        \begin{overpic}[width = \textwidth]{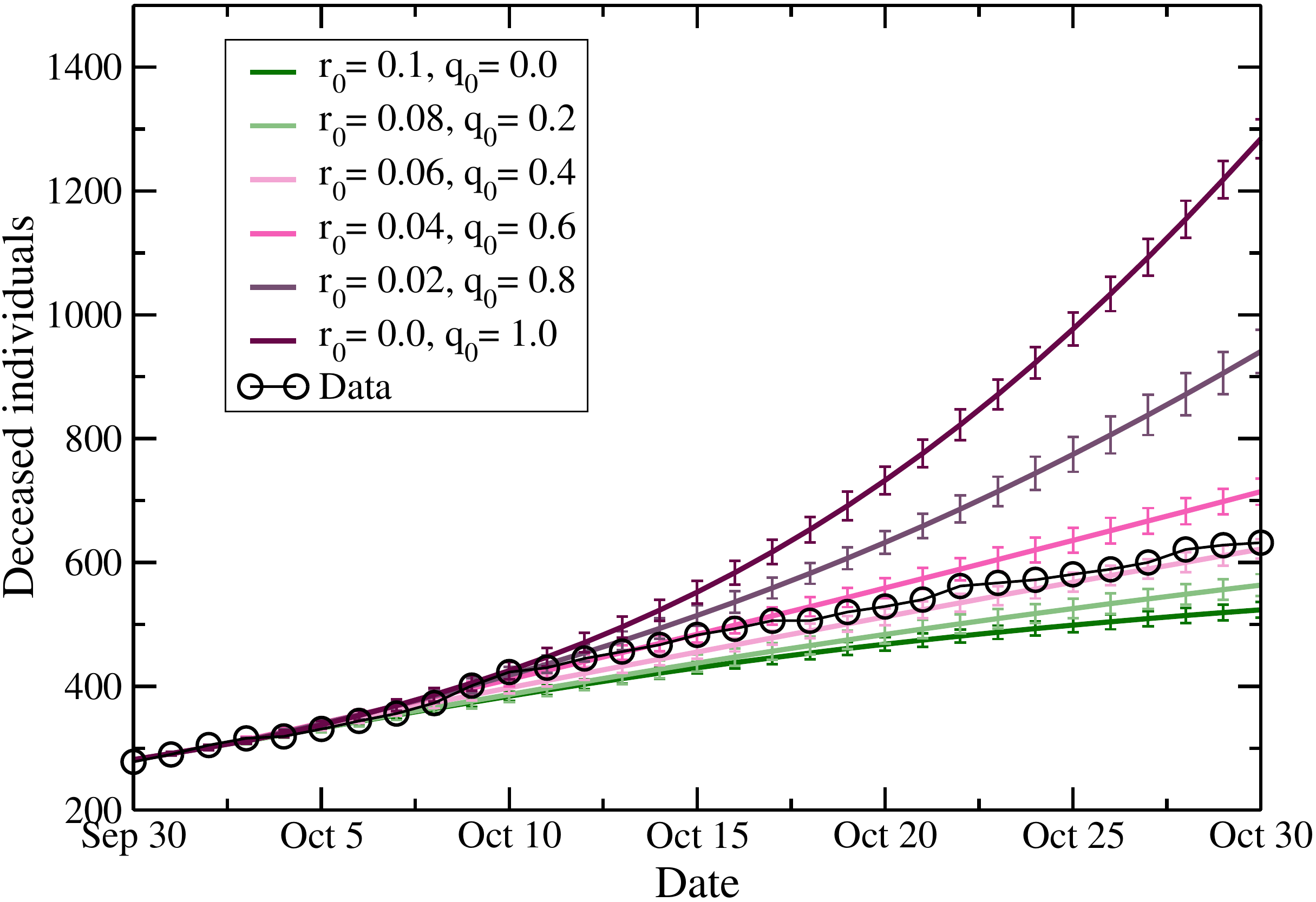}
            \put(75,52){\sffamily \big (B)}
        \end{overpic}
    \end{subfigure}
\caption{\label{data}
{\bf {Comparison with real data.}} {Predictions of our model for the number of infected individuals (panel A) and deaths (panel B) from September 30, 2020 to October 30, 2020, for some values of $r_0$ and $q_0$ (in solid colored lines) \text{*}. In both panels, we also show the actual time series (symbols). The error bars were estimated from 50 stochastic simulations. From these curves, we observe that in a scenario where $q_0\approx 0.4$ (i.e., 40\% of the population is permanently active) and $r_0 \approx 0.06$, the strategies lead to a number of cases and deaths similar to those observed in the real data. However, unlike the real case in October (according to Google’s mobility data \cite{googleMob}), where the mobility was 40\% and 50\% below baseline at workplaces and retail establishments, respectively, under this strategy, $40\%+30\%=70\%$ of the population would be active every week, which may benefit the local economy.}
}
\centering
\end{figure}

\let\thefootnote\relax\footnotetext{\text{*}: We excluded the data point for October 1, 2020, because health authorities in MDP reported a sharp increase in deaths caused by delayed reporting.}

\clearpage

\begin{figure}
\begin{center}
\includegraphics[width = 10cm]{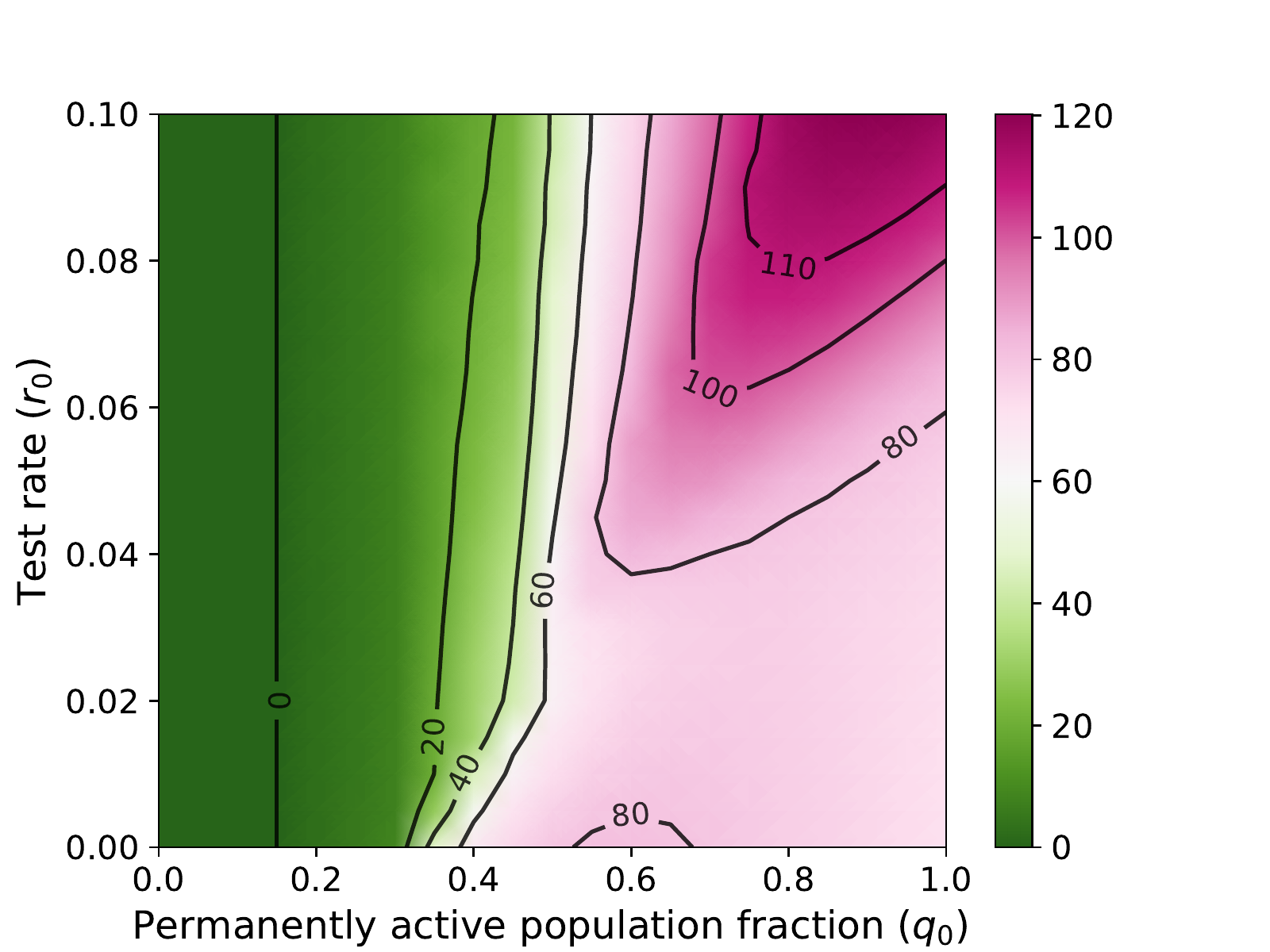}
\end{center}
\vspace*{5mm}
\caption{\label{overw}
{\bf Number of days with an overwhelmed health system.} Heat map of the number of days in which hospitals are overwhelmed in the plane $r_{0}-q_0$ when the number of ICU beds is 60. In the upper right region, this period increases because a higher testing rate delays the extinction of the disease by several weeks, and a weak alternating quarantine ($q_0\approx 1$) leaves a large proportion of the population exposed to COVID-19. 
}
\end{figure}

\clearpage

\paragraph*{S1 Calculation.}
\label{S1_Calculation}
{\bf Inference of contact matrices.} To compute the matrix of an arbitrary setting $\Delta$ (e.g., workplaces), we assume a homogeneous mixing for the contacts occurring at a given location $U$ (e.g., a particular school, company, or household). We denote $a_j^{(k)}$ as the number of individuals of age $j$ interacting with an individual $k$ at a given place $U$ and denote $v_{\Delta}^{(k)}$ as the total number of individuals of all ages within this location. Following \cite{fumanelli2012inferring}, the frequency of contacts between individuals of age groups $i$ and $j$, in all settings of type $\Delta$, is given by
\begin{equation*}
  f^{\Delta}_{ij} = \begin{cases}
    \frac{1}{n_i^{\Delta}} \sum\limits_{k=1}^{N_i}\frac{a^{(k)}_j-\delta_{ij}}{v^{(k)}_{\Delta}-1} &\text{if $v^k_{\Delta}>1, n_i^{\Delta}>0$}\\
    0 &\text{otherwise}
  \end{cases},
\end{equation*}
where $N_i$ is the total number of individuals of age group $i$ in the
entire population, and $n_i^{\Delta}$ is the number of individuals of
age group $i$ with nonzero contacts in the setting $\Delta$. Note that the
Kronecker delta ($\delta_{ij}$) in the numerator avoids counting $k$
as a self-contact and the factor $n_{i}^{\Delta}$ is necessary to normalize its rows, i.e., $\sum_i f_{ij}^{\Delta}=1$. After computing $f^{\Delta}_{ij}$, the contact
matrix associated with setting $\Delta$ is given by
\begin{equation}
 C^{\Delta}_{ij} = \frac{n_i^{\Delta}}{N_i} f^{\Delta}_{ij},
\end{equation}
{where we multiply the frequency of contacts by the probability of having at least one contact in that setting.} Performing this calculation for households, workplaces, schools and
the general community, we obtain a final expression for the contact
matrix by taking the weighted sum
\begin{equation}\label{rrr}
 C_{ij} = \sum\limits_{\Delta} w^{\Delta} \frac{C^{\Delta}_{ij}}{\sum\limits_{i} \sum\limits_{j} C^{\Delta}_{ij}}\ ,
\end{equation}
where the weight $\omega^{\Delta}$ represents the proportion of
transmission events that take place in setting $\Delta$. {Note that the matrices for each setting are divided by a normalization factor $\sum\limits_{i} \sum\limits_{j} C^{\Delta}_{ij}$.}

\begin{figure}
    \centering
    \begin{subfigure}[b]{0.475\textwidth}
        \centering
        \includegraphics[width=\textwidth]{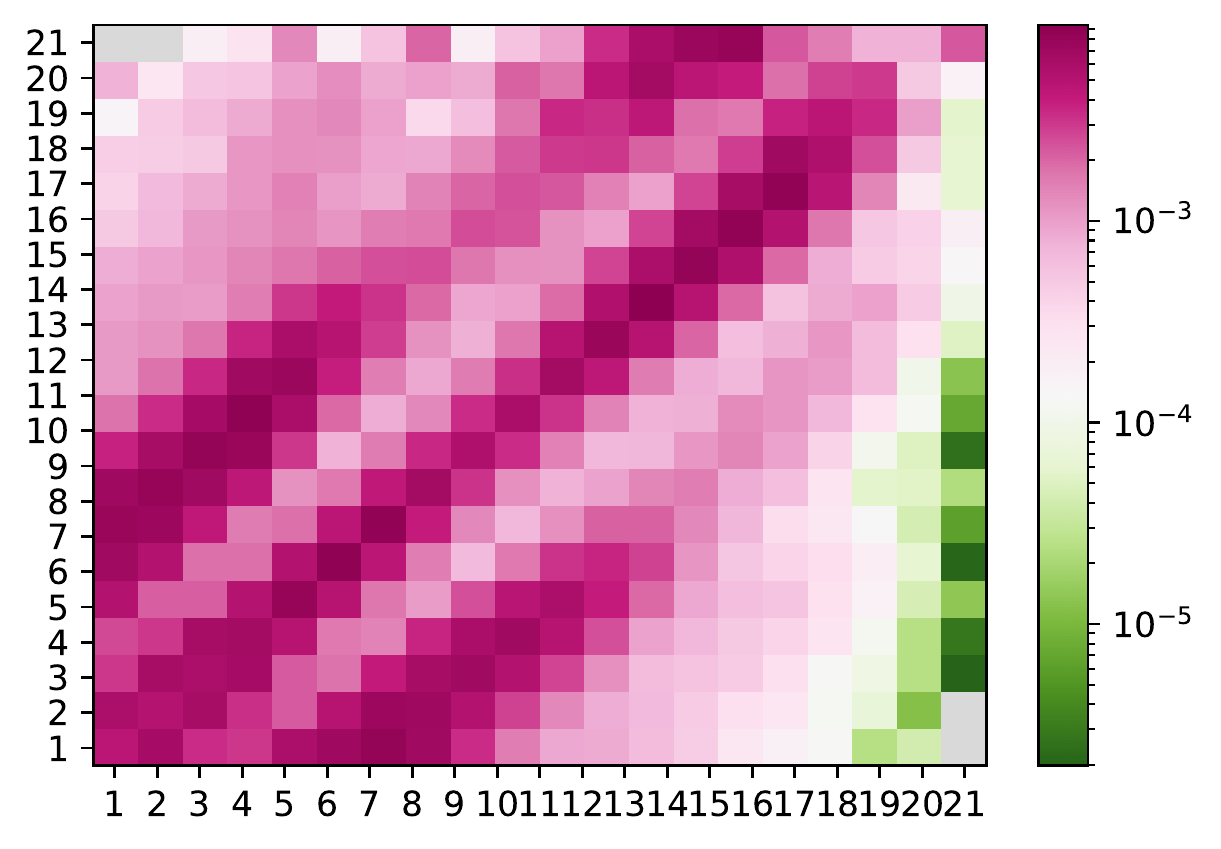}
        \caption{Households}
    \end{subfigure}
    \hfill
    \begin{subfigure}[b]{0.475\textwidth}  
        \centering 
        \includegraphics[width=\textwidth]{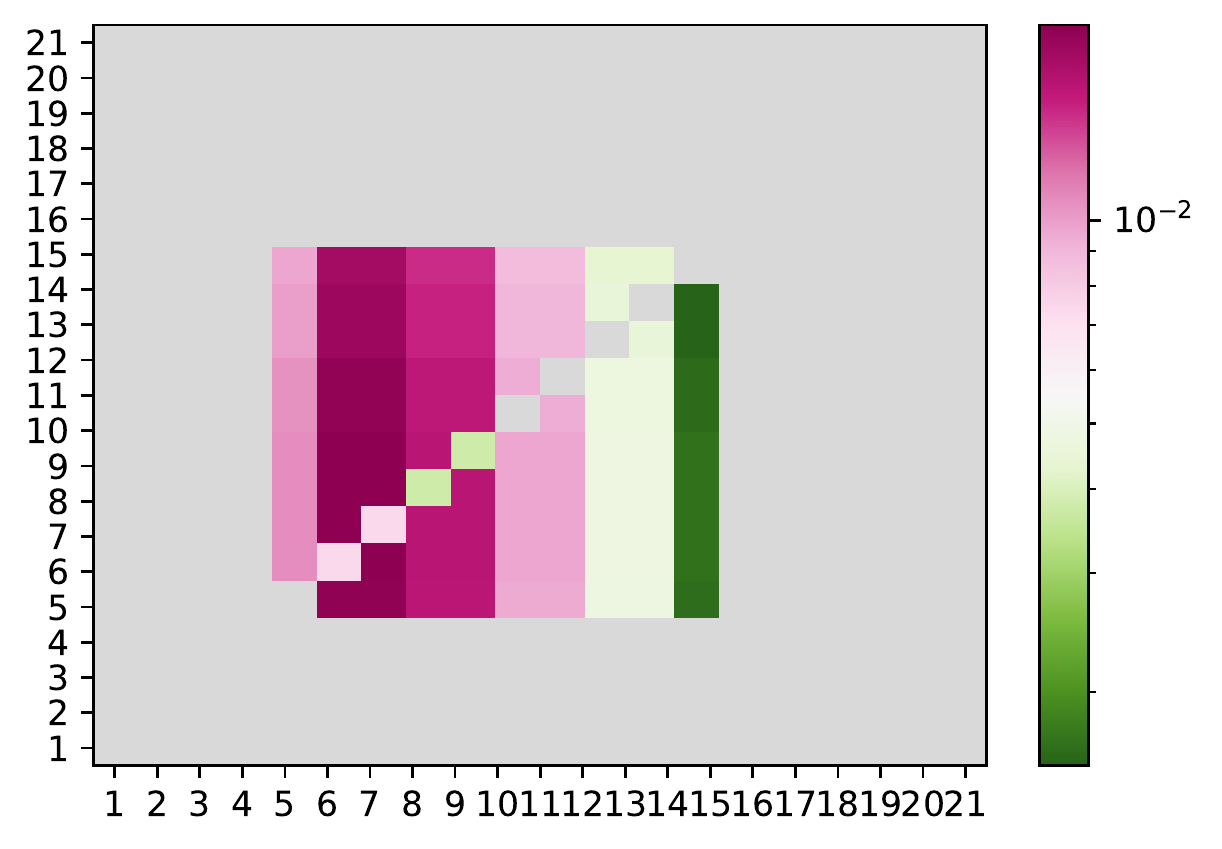}
        \caption{Workplaces}
    \end{subfigure}
    \vskip\baselineskip
    \begin{subfigure}[b]{0.475\textwidth}   
        \centering 
        \includegraphics[width=\textwidth]{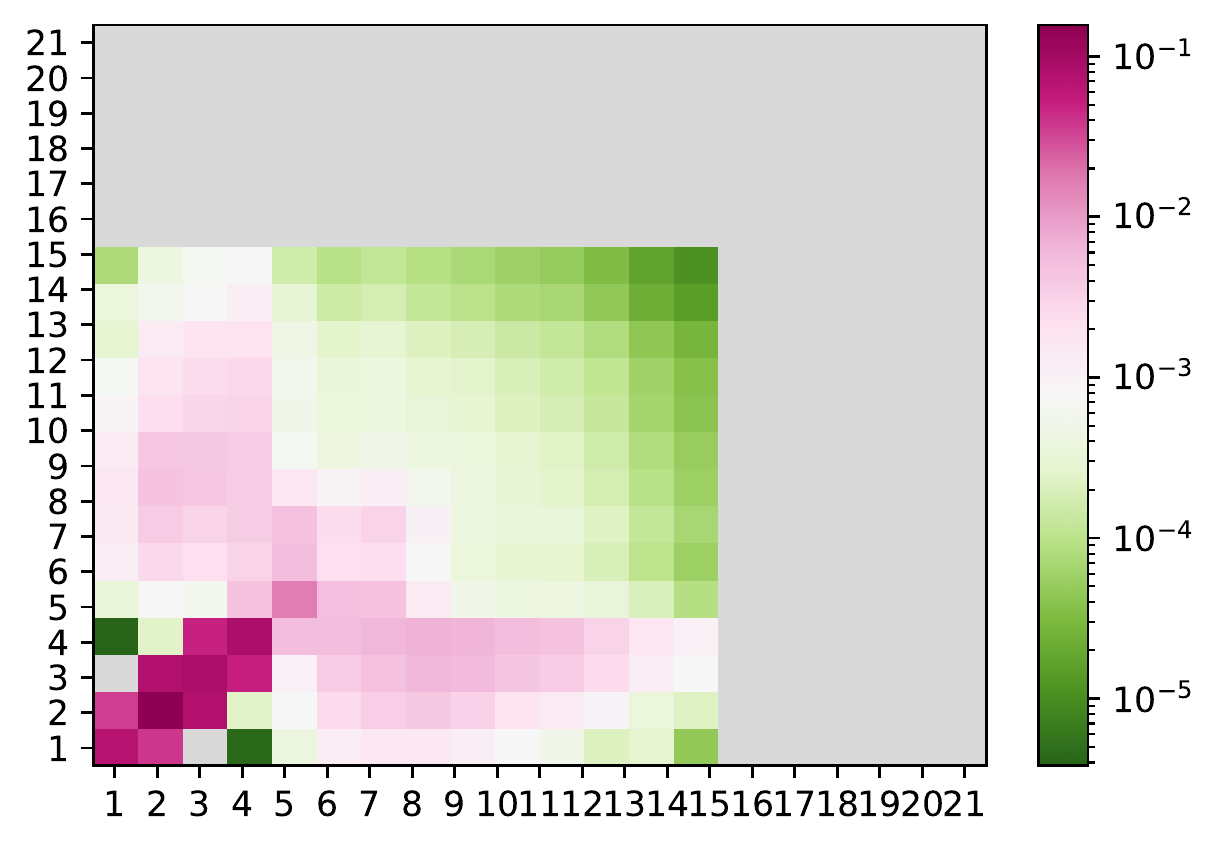}
        \caption{Schools}
    \end{subfigure}
    \hfill
    \begin{subfigure}[b]{0.475\textwidth}   
        \centering 
        \includegraphics[width=\textwidth]{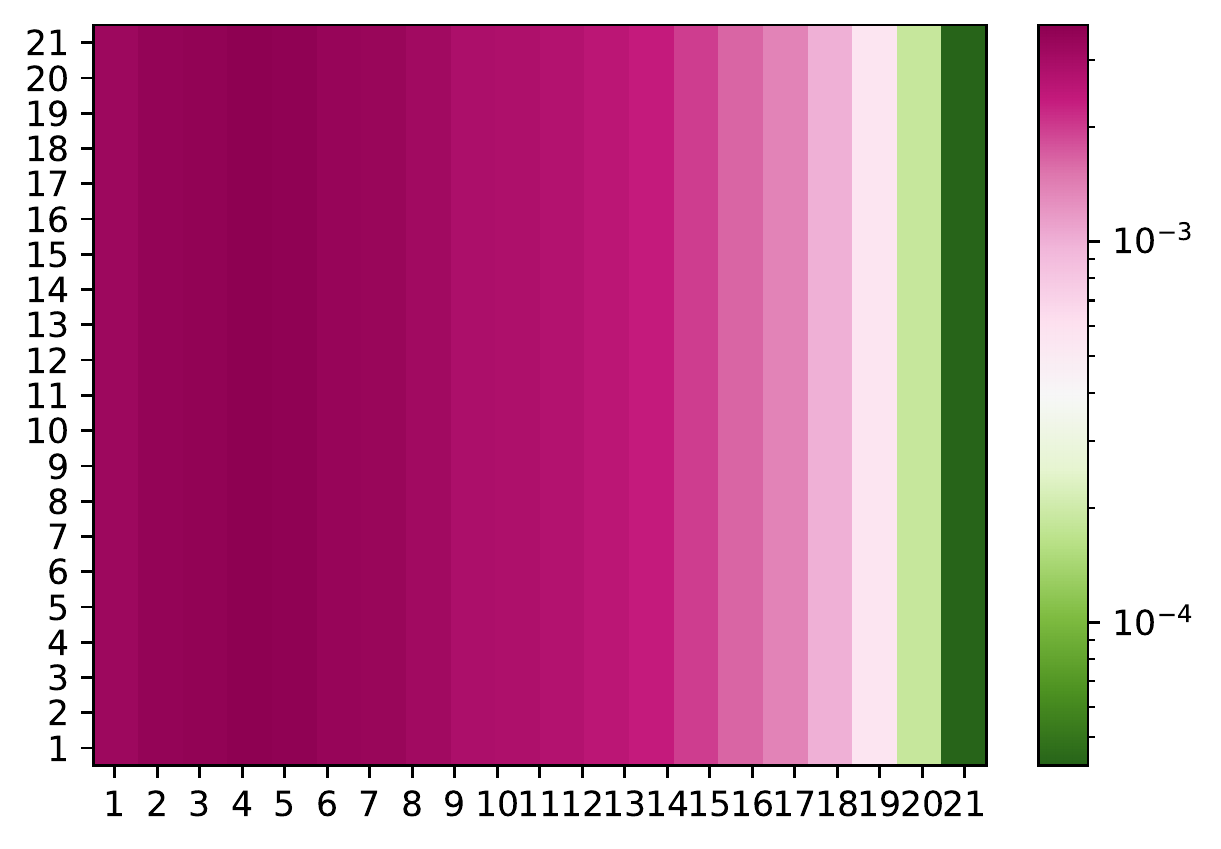}
        \caption{Community}
    \end{subfigure}
\vspace*{5mm}
\caption{\textbf{Contact matrices}. Heat map in logarithmic scale for the contact matrix of the different social settings (see Eq.~(18)). The axes indicate the 21 age cohorts considered in our model, each covering an age range of 5 years. For example, the first cohort covers from 0 to 4 years old, the second one from 5 to 9 years old, and so on. The gray color corresponds to $C_{ij}^{\Delta}=0$. \label{mat_settings}}
\end{figure}

{Fig. \ref{mat_settings} includes a graphic representation of the normalized matrices $\frac{C^{\Delta}_{ij}}{\sum\limits_{i} \sum\limits_{j} C^{\Delta}_{ij}}$ that were calculated for the four different settings\text{*}. 
\let\thefootnote\relax\footnotetext{\text{*}: We have made the contact matrices publicly available on GitHub \cite{github}.}
It can be seen that the school and work matrices concentrate contacts in specific regions, while the home matrix is more dispersed. Regarding the contact matrix within workplaces, the low values on the diagonal are due to lack of sufficient data to compute this matrix. On the other hand, the community contact matrix represents contacts that are totally random, so that the frequency of contacts with a certain age group is proportional to the population in this group.} 

\renewcommand{\thesubfigure}{\Alph{subfigure}}

\begin{figure}
    \centering
    \begin{subfigure}[b]{0.475\textwidth}
        \centering
        \includegraphics[width=\textwidth]{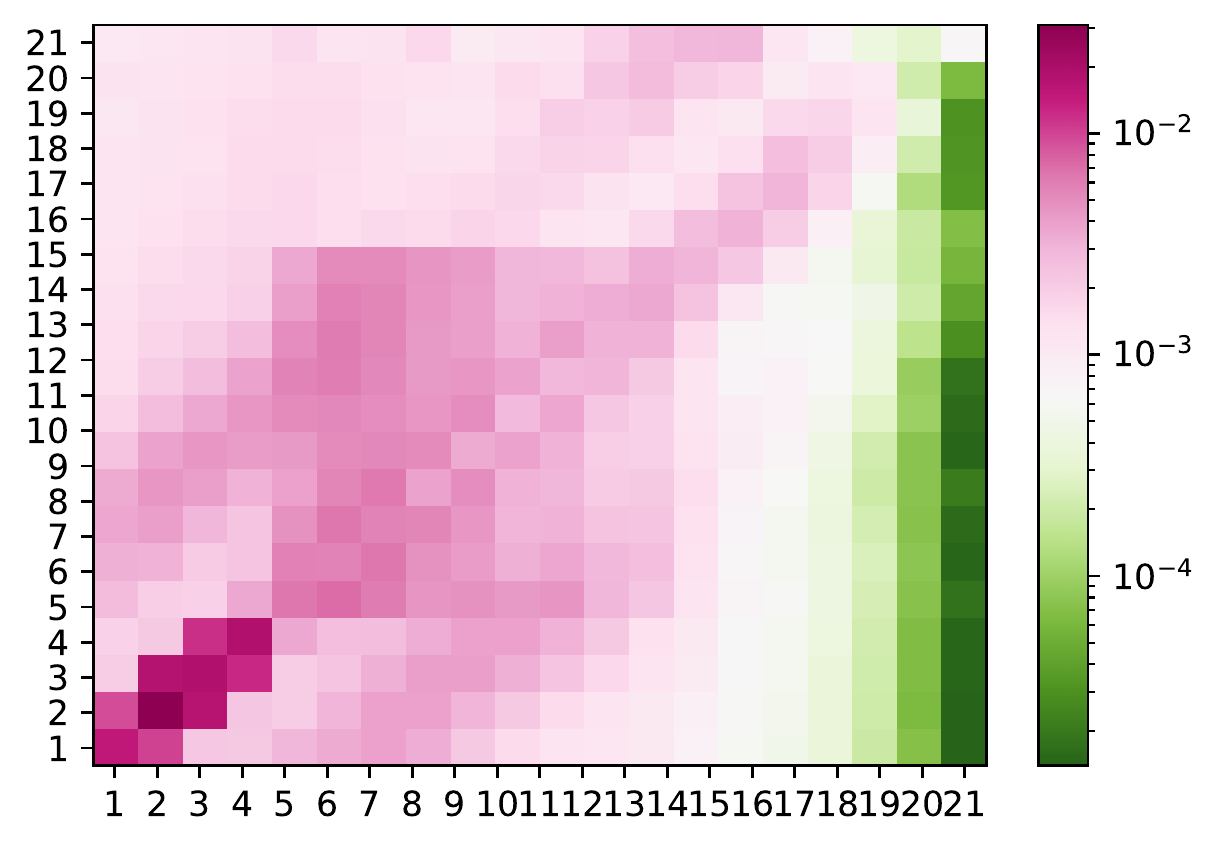}
        \caption{Non-quarantined population}
    \end{subfigure}
    \hfill
    \begin{subfigure}[b]{0.475\textwidth}   
        \centering 
        \includegraphics[width=\textwidth]{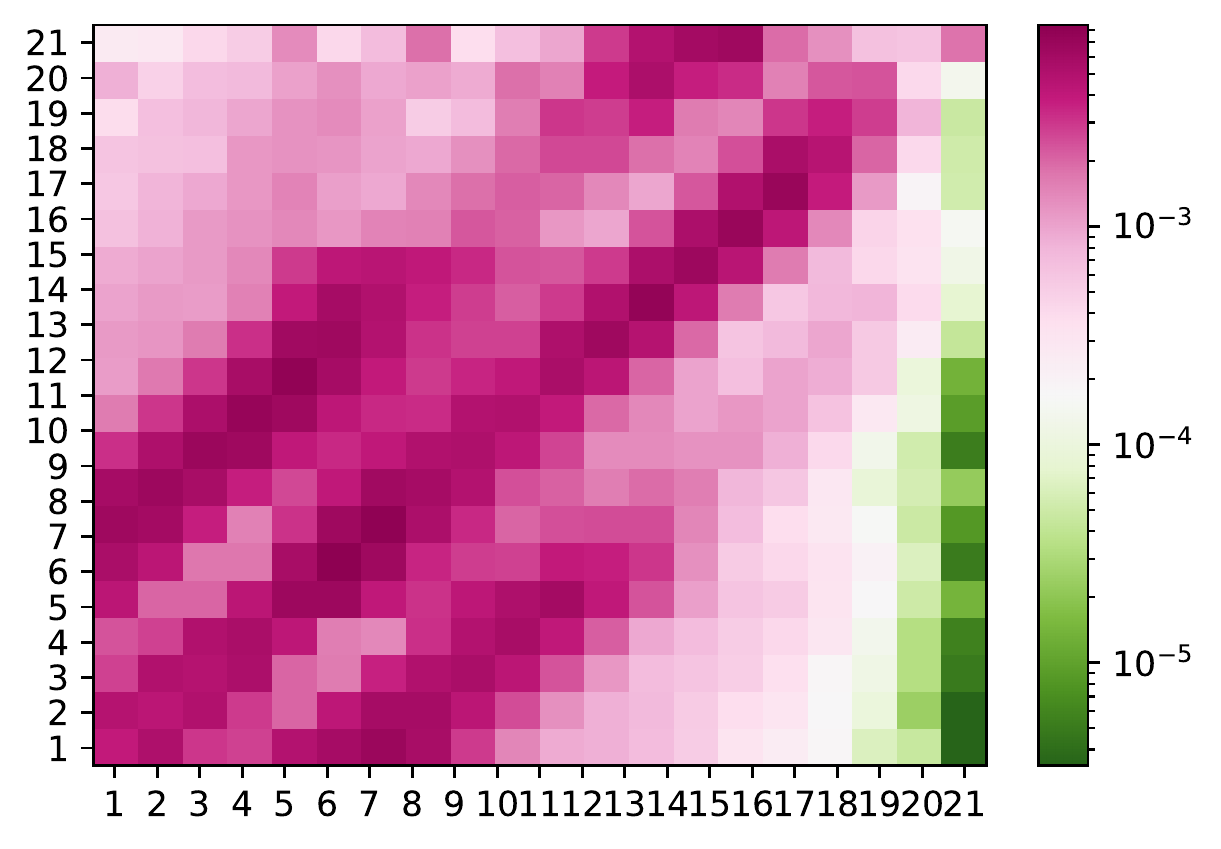}
        \caption{August 18 to September 30, 2020}
    \end{subfigure}
\vspace*{5mm}
\caption{\textbf{Total contact matrices}. Heat map in logarithmic scale for the contact matrix given by Eq.~(19) for: non-quarantined individuals in the AQ strategy (panel A) and for the period August 18 to September 30, 2020 (panel B). \label{mat_total}}
\end{figure}

In \cite{fumanelli2012inferring}, the authors estimate the following weights based on empirical data for influenza-like diseases:
\begin{align*}
  w^{house} &= 0.30,\\
  w^{work} &= 0.19,\\
  w^{school} &= 0.18,\\
  w^{comm} &= 0.33.
\end{align*}
We use these weight's values in Eq. (19) to fit $\beta_\textnormal{free}$ (see Sec. 3.1) and for the non-quarantined population in the AQ strategy (see Fig. \ref{mat_total}A). According to Google mobility reports as of August 14 (average of 7 days), activity in workplaces was decreased by 31\%, and in transit stations and retail \& recreation by 75\% (average between both settings). Added to the fact that all schools were closed, we used these percentages to decrease the previous weight factors, which we used to fit the curve of accumulated deaths in General Pueyrred\'on from August 18 to September 30 (prior to implementing the mitigation strategies in our simulations):
\begin{align*}
  w^{house} &= 0.79,\\
  w^{work} &= 0.13,\\
  {w^{school}} &= {0.0},\\
  {w^{comm}} &= {0.08}.
\end{align*}
In this way, we estimate the number of individuals in each compartment segregated by age group, on September 30, 2020 (see Sec. 3.1 and Fig. \ref{mat_total}B).

\clearpage
{
\paragraph*{S2 Calculation.}
\label{S2_Calculation}
{\bf Simplified analysis of the effect of a threshold in the testing strategy on the accumulated number of infected cases.} In Sec. 2.4 we included in our model a testing rate that obeys the following function:
\begin{eqnarray}\label{eqAp.rtest}
   r_{test}(t)=r_0 \; \Theta(I(t)-I_0),
\end{eqnarray}
where $I_0$ is a threshold for the number of infected individuals, above which random testing is conducted with a rate $r_{test}=r_0$, whereas below $I_0$, the testing strategy is suspended ($r_{test}=0$). In Fig. \ref{massResult}B it was observed that the strategy cannot contain the number of accumulated cases $I_{ac}$ from the start of the epidemic, even for large values of $r_0$. This is because the testing strategy is suspended when $I<I_0$, regardless of the value of $r_0$. To determine analytically the final number of accumulated cases for $r_0\to \infty$, in this secion we present a susceptible-infected-removed model without age structure, where individuals are tested at a rate $r_{test}$ given by Eq.~(\ref{eqAp.rtest}). For this model, the dynamic equations are as follows:
\begin{eqnarray}
\frac{dS}{dt}&=&-\beta\frac{S}{N}I,\label{eq.ApSiT1}\\
\frac{dI}{dt}&=&\beta\frac{S}{N}I-(\gamma+r_{test}(t))I,\\
\frac{dR}{dt}&=&(\gamma+r_{test}(t))I,\label{eq.ApSiT3}
\end{eqnarray}
where:
\begin{enumerate}
\item $S$, $I$, and $R=N-S-I$ are the number of susceptible, infected, and removed individuals, respectively, and $N$ denotes the size of the population. Here an individual is removed if he/she recovers or tests positive.
\item $\beta$ is the infection rate.
\item $\gamma$ is the recovery rate.
\end{enumerate}
When the strategy is applied, i.e. $I>I_0$, the effective reproduction number at time $t$ \cite{ehrhardt2019sir} is
\begin{eqnarray}\label{eqapRon}
R_t^{on}=\frac{\beta}{(\gamma+r_0) N}S(t),
\end{eqnarray}
where the superscript "on" indicates that the strategy is implemented. Analogously, when $I<I_0$, the effective reproduction number is
\begin{eqnarray}\label{eqapRoff}
R_t^{off}=\frac{\beta}{\gamma N}S(t).
\end{eqnarray}
Note that $R_t^{off}\geq R_t^{on}$.
In Fig. \ref{fig.Apbet02}A) we show the time evolution of the fraction of removed individuals, where it is observed (as in Fig. \ref{massResult}B) that the effect of an increasing value of $r_0$ on $R(t\to \infty)$ becomes negligible. On the other hand, in Fig. \ref{fig.Apbet02}B) we show the time evolution of active cases. After this magnitude reaches the threshold of $I_0/N$, $I(t)/N$ oscillates around $I_0/N$ for a period of time because the testing strategy is applied and suspended intermittently when $I(t)$ is greater and less than $I_0$, respectively. In addition, we observed that the oscillation amplitude is small compared with $I_0$ (see inset of Fig. \ref{fig.Apbet02}B), so the infection curve flattens or becomes approximately a plateau. However, during this period, the number of susceptible individuals is still decreasing (see Fig. \ref{fig.Apbet02}C), and the curve $I(t)$ remains approximately flattened until the number of susceptible individuals reaches a value such that $R_t^{off}= 1$, that is, the number of cases cannot increase even when the strategy is suspended. From this moment, that we call $t=t^*$, the population has achieved herd immunity. Using that $r_0\gg 1$ and $I(t^*)\approx I_0$, the size of the susceptible population at time $t=t^*$, denoted as $S^*$, is given by the condition $R_t^{off}=1$, i.e.,
\begin{eqnarray}\label{eq.Sstar}
  R_{t^*}^{off}=1 \Rightarrow \frac{\beta}{\gamma N}S^* =1  \Rightarrow S^* =\frac{\gamma N}{\beta}.
\end{eqnarray}
Then, for $t\geq t^*$, the number of infected individuals remains below $I_0$, and Eqs.~(\ref{eq.ApSiT1})-(\ref{eq.ApSiT3}) are reduced to an SIR model without testing intervention, which is described by the following set of equations:
\begin{eqnarray}
\frac{dS}{dt}&=&-\beta\frac{S}{N}I,\label{eq.ApNoT1}\\
\frac{dI}{dt}&=&\beta\frac{S}{N}I-\gamma I,\\
\frac{dR}{dt}&=&\gamma I,\label{eq.ApNoT3}
\end{eqnarray}
If we assume that $r_0\gg 1$, the initial condition for these equations at $t=t^*$, are given by: $S(t^*)=S^*$, $I(t^*)=I_0$, and $R(t^*)=N-S(t^*)-I(t^*)$, and the solution of Eqs.~(\ref{eq.ApNoT1})-(\ref{eq.ApNoT3}) satisfies the following equation:
\begin{eqnarray}\label{eq.casimir}
S(t^*)+I(t^*)-\frac{\gamma N}{\beta}\ln(S(t^*))=S(t)+I(t)-\frac{\gamma N}{\beta}\ln(S(t)),
\end{eqnarray}
for $t>t^{*}$. Finally, at infinite time, as $I(t)\to 0$, and $S(t \to \infty)=N-R(t \to \infty)$, the Eq.~(\ref{eq.casimir}) is reduced to:
\begin{eqnarray}\label{eq.casimirtinfR}
S(t^*)+I(t^*)-\frac{\gamma N}{\beta}\ln(S(t^*))=(N-R(t \to \infty))-\frac{\gamma N}{\beta}\ln(N-R(t \to \infty)),
\end{eqnarray}
and using that the number of accumulated cases $I_{ac}$ is equal to the number of removed individuals $R$, in the limit $t\to \infty$, and replacing Eq. (\ref{eq.Sstar}) and $I(t^*)=I_0$ in Eq. (\ref{eq.casimirtinfR}), we finally obtain that $I_{ac}(t\to \infty)$ for $r_0 \gg 1$ satisfies the equation:
\begin{eqnarray}\label{eq.casimirtinfIacumm}
\frac{\gamma N}{\beta}+I_0-\frac{\gamma N}{\beta}\ln(\frac{\gamma N}{\beta})=(N-I_{ac}(t \to \infty))-\frac{\gamma N}{\beta}\ln(N-I_{ac}(t \to \infty)).
\end{eqnarray}
The solution of $I_{ac}(t\to \infty)$ is shown as a dashed line in Fig. \ref{fig.Apbet02}A), which agrees to the limit number of accumulated cases in the final state as $r_0$ increases. Therefore, as mentioned in Sec. 3.2, a testing strategy with a test rate given by Eq.~(\ref{eqAp.rtest}) is effective in flattening the curve of infections but cannot completely contain the disease spread from the start of the epidemic.

\begin{figure}
\vspace{0.5cm}
\begin{center}
\includegraphics[scale=0.35]{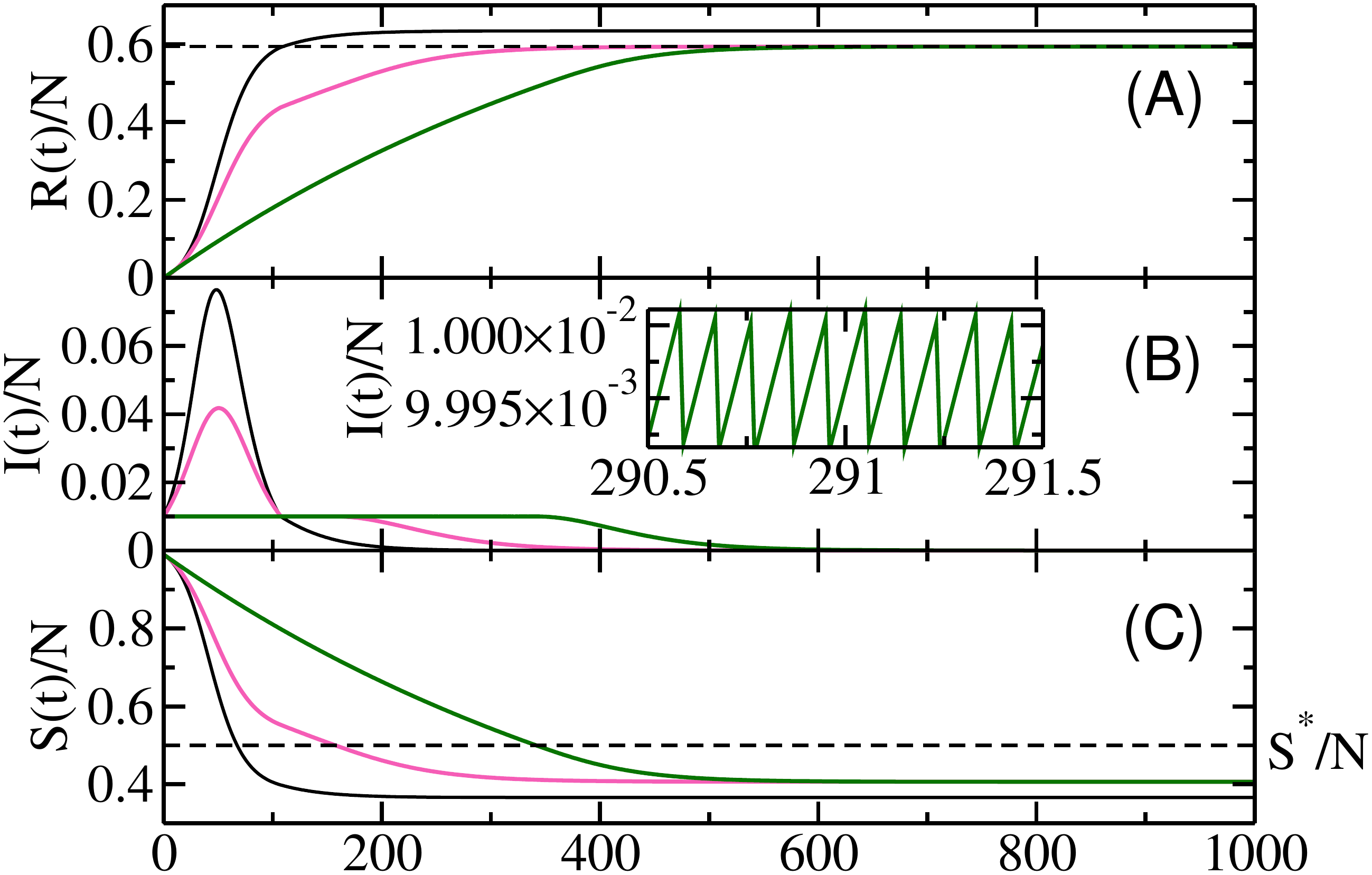}
\end{center}
\caption{{\textbf{Simplified model results}. Time evolution of the fraction of accumulated cases (panel A), active cases (panel B), and susceptible individuals (panel C) obtained from Eqs.~(\ref{eq.ApSiT1})-(\ref{eq.ApSiT3}) with $\beta=0.2$, $\gamma=0.1$, $I_0=0.01$, and different values of $r_0$: 0.03 (solid black), 0.05 (solid pink), and 0.10 (solid green). Here, we also use as initial condition $I(t=0)=I_0$, and $S(t=0)=N-I_0$. In panel (A), the dashed line indicates the final number of accumulated cases for $r_0\gg 1$, predicted by Eq.~(\ref{eq.casimirtinfIacumm}). In panel (B) the inset shows an enlargement of the main figure. In panel (C), the dashed line indicates the value of $S^*$, obtained from Eq.~(\ref{eq.Sstar}).}}\label{fig.Apbet02}
\end{figure}

}

\clearpage
\paragraph*{S1 Table.}
\label{S1_Table}
{\bf Age-independent parameters.} Table describing the age-independent parameters and their values.

\begin{table}[!ht]
\small
\centering
\begin{tabular}{|l|l|l|l|l|}
\hline
Symbol & Description & Value {[$days^{-1}$]} & Reference & Notes \\ \hline
$\beta_\textnormal{free}$ & rate $S \to I$ & 5.6 & \cite{wu2020nowcasting, li2020early, riou2020pattern} & \dag\\
\hline
$\alpha$ & rate $E \to I$ & 1/(5.1-2.0) & \cite{lauer2020incubation, linton2020incubation, li2020early, Byrnee039856} & \ddag\\
\hline
$\omega$ & rate $I \to M, \mathcal{H}$ & 1/(4.0+2.0) & \cite{datosabiertos, Byrnee039856} & \ddag\\ 
\hline
$\gamma$ & rate $H \to R$ & 1/8 & \cite{ferguson2020report} & \\ \hline
$\gamma^*$ & rate $H^* \to H$ & 1/10 & \cite{ferguson2020report} & \\ \hline
$\gamma^m$ & rate $M \to R$ & 1/14 & \cite{aislamiento_milds} & \textcurrency\\ \hline
$\gamma^u$ & rate $U \to R$ & 1/9.5 & \cite{hu2020clinical} & \\ \hline
$\delta$ & rate $H^\dagger \to D$ & 1/8 & \cite{datosabiertos} & \\ \hline
$\delta^*$ & rate $H^{\dagger*} \to D$ & 1/11 & \cite{datosabiertos} & \\ \hline
\end{tabular}
\end{table}

\dag: These works report the value of $R_0$ in the stage of free propagation, from which we calculate $\beta_\textnormal{free}$ using the next generation matrix method \cite{diekmann2010construction}.

\ddag: We subtract two days from the time of transition $E\to I$ reported in \cite{lauer2020incubation, linton2020incubation, li2020early}, since in \cite{Byrnee039856} the authors estimated that patients are infectious two days before showing symptoms. Therefore, infectious patients also stay two additional days in the community, increasing the time of transition $I\to M$ and $I\to \mathcal{H}$ reported in \cite{datosabiertos}.

\textcurrency: Health authorities in Argentina require mild patients to remain self-isolated at home, and these cases are reported as recovered after 14 days since their first symptoms appeared.

\clearpage
\paragraph*{S2 Table.}
\label{S2_Table}
{\bf Age-dependent parameters.} Table describing the age-dependent parameters and their values. The age groups are 5 years wide.

\begin{table}[!ht]
\small
\centering
\begin{tabular}{|l|p{3cm}|p{7.8cm}|l|}
\hline
Symbol & Description & Values for Ages [0-4, 5-9, ..., 100-104] & Ref. \\ \hline
{$N_j$} & {Number of individuals in age group $j$} & {44797, 47945, 48457, 51013, 50015, 47266, 46560, 43238, 39557, 39067, 37358, 34966, 32699, 27311, 22564, 18999, 13780, 7680, 2532, 582, 58}
 & \cite{censo} \\ \hline
$\epsilon$ & Proportion of infected individuals that will develop symptoms & 1/3, 1/3, 1/3, ..., 1/3, 1, 1, 1 & \cite{quiros_prevalence, quiros_prevalence2}\textsuperscript{\dag} \\ \hline
$\zeta$ & Proportion of cases that need hospitalization ($\mathcal{H}$) & 0.07504484, 0.03803735, 0.02837104, 0.0211873, 0.01936113, 0.02174475, 0.02556259, 0.02665036, 0.03077249, 0.03761504, 0.04842162, 0.0635869, 0.09234972, 0.12900137, 0.17795732, 0.22363486, 0.2618449, 0.28956643, 0.29179373, 0.28364116, 0.1965602 &\cite{datosabiertos}\ddag \\
\hline
$\theta$ & Proportion of hospitalized patients in general beds who will fully recover & 0.93138011, 0.94563427, 0.96491228, 0.9447619, 0.95835351, 0.95063985, 0.94168139, 0.91165612, 0.86417323, 0.81459711, 0.73918075, 0.65684799, 0.53099676, 0.42928786, 0.37042989, 0.31017208, 0.29837518, 0.25360883, 0.25273159, 0.25426357, 0.1625 & \cite{datosabiertos}\ddag \\
\hline
$\theta^*$ & Proportion of hospitalized patients in ICUs who will fully recover & 0.04240555, 0.04448105, 0.02232855, 0.03047619, 0.01985472, 0.019805, 0.02726584, 0.03718354, 0.05068898, 0.05412687, 0.06202268, 0.07073748, 0.06057536, 0.05436309, 0.04901567, 0.03942496, 0.03126539, 0.02462496, 0.02185273, 0.01705426, 0.0 & \cite{datosabiertos}\ddag \\
\hline
$\theta^\dagger$ & Proportion of hospitalized patients in general beds who will die & 0.00693909, 0.00329489, 0.00637959, 0.01142857, 0.00920097, 0.01279707, 0.01514769, 0.02109705, 0.03617126, 0.05167769, 0.08585975, 0.11180391, 0.18314425, 0.26018054, 0.34491764, 0.42191244, 0.48375185, 0.55165582, 0.59619952, 0.62325581, 0.65 & \cite{datosabiertos}\ddag \\ 
\hline
$\theta^{\dagger*}$ & Proportion of hospitalized patients in ICUs who will die & 0.01927525, 0.00658979, 0.00637959, 0.01333333, 0.0125908, 0.01675807, 0.01590507, 0.03006329, 0.04896654, 0.07959833, 0.11293682, 0.16061062, 0.22528363, 0.25616851, 0.2356368, 0.22849052, 0.18660758, 0.17011039, 0.12921615, 0.10542636, 0.1875 & \cite{datosabiertos}\ddag \\ 
\hline
\end{tabular}
\end{table}
\FloatBarrier

\dag: This value was obtained from a seroprevalence study in the city of Buenos Aires, Argentina~ \cite{quiros_prevalence, quiros_prevalence2}.

{\ddag: \cite{datosabiertos} is a detailed dataset of all officially reported COVID-19 cases in Argentine health institutions, including the type of treatment and its outcome.}

\clearpage
{
\paragraph*{S3 Table.}
\label{S3_Table}
{\bf Estimation of initial values.} This table shows the estimated population in the different compartments by age groups as of September 30, 2020, which are used as initial conditions for the AQ and massive testing strategies.

\begin{table}[!ht]
{
\small
\centering
\begin{tabular}{|l|p{9cm}|l|}
\hline
Compartment & Values for Ages [0-4, 5-9, ..., 100-104] & Total \\ \hline
$S$ & 39469.7, 42322.4, 42827.7, 45090.7, 42681.2, 40441.2, 39943.5, 37147.8, 34024.8, 33640.0, 32117.3, 30080.9, 28367.3, 23923.8, 20704.4, 17545.8, 12763.4, 7129.8, 2332.7, 537.9, 52.5 & 573144.6 \\ \hline
$E$ & 472.0, 498.9, 500.0, 525.8, 640.6, 596.8, 579.7, 534.5, 485.9, 476.5, 459.4, 428.2, 381.2, 299.9, 168.7, 132.4, 92.67, 50.13, 18.12, 4.034, 0.5280 & 7346.0 \\ \hline
$U$ & 771.8, 815.3, 816.7, 859.0, 1054.1, 981.5, 952.6, 877.7, 797.6, 782.3, 754.7, 703.5, 625.2, 490.6, 273.1, 213.9, 149.7, 80.99, 0.000, 0.000, 0.000 & 12000.2 \\ \hline
$I$ & 264.3, 279.2, 279.7, 294.2, 360.3, 335.6, 325.7, 300.2, 272.8, 267.6, 258.1, 240.6, 213.9, 167.9, 93.76, 73.48, 51.41, 27.82, 30.21, 6.720, 0.8818 & 4144.3 \\ \hline
$M$ & 406.0, 445.8, 451.0, 477.9, 590.8, 548.6, 530.0, 487.5, 441.0, 429.6, 410.0, 376.1, 323.5, 243.1, 126.6, 93.53, 62.22, 32.41, 35.16, 7.900, 1.168 & 6519.9 \\ \hline
$H$ & 20.84, 11.34, 8.488, 6.573, 7.439, 7.716, 8.770, 8.225, 8.289, 9.423, 10.76, 11.90, 12.46, 11.13, 7.358, 6.036, 4.683, 2.370, 2.568, 0.5504, 0.03054 & 166.9 \\ \hline
$H^*$ & 1.093, 0.6133, 0.2300, 0.2466, 0.1809, 0.1887, 0.2962, 0.3879, 0.5547, 0.7102, 1.011, 1.412, 1.559, 1.531, 1.053, 0.8326, 0.5409, 0.2550, 0.2480, 0.04181, 0.000 & 12.98 \\ \hline
$H^\dagger$ & 0.1500, 0.0381, 0.05514, 0.07760, 0.07030, 0.1022, 0.1380, 0.1846, 0.3320, 0.5687, 1.174, 1.871, 3.952, 6.146, 6.224, 7.486, 7.032, 4.800, 5.684, 1.284, 0.1222 & 47.49 \\ \hline
$H^{\dagger*}$ & 0.5337, 0.09765, 0.07062, 0.1159, 0.1233, 0.1716, 0.1858, 0.3371, 0.5760, 1.123, 1.979, 3.445, 6.230, 7.752, 5.437, 5.182, 3.468, 1.892, 1.575, 0.2776, 0.04508 & 40.62 \\ \hline
$R$ & 3389.5, 3571.4, 3573.3, 3758.3, 4680.3, 4354.0, 4218.4, 3880.5, 3523.2, 3454.8, 3334.2, 3102.7, 2732.5, 2116.5, 1141.3, 880.0, 610.8, 327.6, 81.47, 18.18, 2.521 & 52751.4 \\ \hline
$D$ & 1.981, 0.4014, 0.3919, 0.5958, 0.5957, 0.8447, 1.015, 1.594, 2.784, 5.136, 9.704, 16.25, 31.42, 43.47, 36.91, 40.70, 34.56, 22.35, 24.81, 5.387, 0.5658 & 281.5 \\ \hline
\hline
\end{tabular}
}
\end{table}
\FloatBarrier
}

\clearpage
\paragraph*{Impact of $I_0$ on the total number of tests.} In \nameref{S4_Table} and \nameref{S5_Table} we show the total number of tests used and the total number of cases predicted as of June 7, 2021, when only the testing strategy is implemented ($q_0=1)$, for different values of $r_0$ and $I_0$. As expected, as the $I_0$ threshold is decreased the number of cases decreases but at the cost of a larger number of tests, especially for large values of $r_0$.

\paragraph*{S4 Table.}
\label{S4_Table}
{\bf Total number of tests.}

\begin{table}[!ht]
\small
\centering
\begin{tabular}{|l|l|l|l|l|}
\hline
$r_0$ ($q_0=1$) & $I_0=1$ & $I_0=2072$ & $I_0=4144$ & $I_0=8288$ \\ \hline
0.02 & 809857 & 536180 & 504002 & 352928\\
\hline
0.04 & 1898939 & 1135992 & 1044940 & 681154\\
\hline
0.06 & 3342937 & 1791866 & 1591372 & 950400\\
\hline
0.08 & 5208377 & 2484934 & 2121557 & 1158639\\
\hline
0.10 & 7977256 & 3342067 & 2701127 & 1321335\\
\hline

\end{tabular}
\end{table}

\paragraph*{S5 Table.}
\label{S5_Table}
{\bf Total number of cases.}

\begin{table}[!ht]
\small
\centering
\begin{tabular}{|l|l|l|l|l|}
\hline
$r_0$ ($q_0=1$) & $I_0=1$ & $I_0=2072$ & $I_0=4144$ & $I_0=8288$ \\ \hline
0.02 & 178080 & 178850 & 179586 & 181668\\
\hline
0.04 & 163608 & 166244 & 168707 & 174310\\
\hline
0.06 & 148336 & 155907 & 161738 & 171228\\
\hline
0.08 & 132753 & 152112 & 160737 & 171162\\
\hline
0.10 & 117361 & 152129 & 160845 & 171308\\
\hline

\end{tabular}
\end{table}

\newpage
\section*{Acknowledgments}
The authors wish to acknowledge the statistical office that provided the underlying data making this research possible: National Institute of Statistics and Censuses, Argentina. LV, IAP, MT, LDV, CEL and LAB acknowledge UNMdP (EXA 956/20), CONICET (PIP 00443/2014) and Ministerio de Ciencia, Tecnología e Innovación, Argentina (BUE 193 - {\it Programa de articulación y fortalecimiento federal de las capacidades en ciencia y tecnología COVID-19}) for financial support. We also thank the Physics Department of Boston University, for allowing us to use its facilities. Lastly, we thank the anonymous reviewers for their careful reading of our manuscript and their insightful comments and suggestions.

\bibliographystyle{elsarticle-num}

\end{document}